\documentclass[aps,twocolumn,floatfix,superscriptaddress,nofootinbib]{revtex4-2}

\usepackage{booktabs}
\usepackage{graphicx}
\usepackage{braket}
\usepackage{stix}
\usepackage{xcolor}
\usepackage{amsfonts} 
\usepackage{appendix}
\usepackage{amsmath}
\usepackage{float}

\usepackage{dcolumn}
\usepackage{soul}
\usepackage{bm}

\usepackage{amsthm}
\usepackage{enumitem}

\usepackage{amsmath}

\usepackage{xr-hyper}
\usepackage{hyperref}
\hypersetup{colorlinks,
	citecolor=blue
}
\usepackage{cleveref}
\usepackage{apptools}
\usepackage{svg}
\AtAppendix{\counterwithin{lemma}{section}}
\usepackage{ulem}

\makeatletter
\newcommand*{\addFileDependency}[1]{
  \typeout{(#1)}
  \@addtofilelist{#1}
  \IfFileExists{#1}{}{\typeout{No file #1.}}
}
\makeatother

\usepackage{fancyhdr,graphicx,amsmath,amssymb}
\usepackage{graphicx}
\usepackage{afterpage}
\usepackage[ruled,vlined]{algorithm2e}
\usepackage{tikz}
\usetikzlibrary{quantikz2}
\usepackage{listings}
\include{pythonlisting}

\begin{document}

\preprint{APS/123-QED}

\title{Transmon qutrit-based simulation of spin-1 AKLT systems}

\author{Keerthi Kumaran}

\affiliation{Department of Physics and Astronomy, Purdue University, West Lafayette, IN 47907, USA}
\affiliation{Quantum Science Center, Oak Ridge National Laboratory, Oak Ridge, TN 37831, USA}

\author{Faisal Alam}
\affiliation{Theoretical Division, Los Alamos National Laboratory, Los Alamos, NM 87545, USA}
\affiliation{Department of Physics, University of Illinois at Urbana-Champaign, IL 61801, USA}

\author{Norhan Eassa}

\affiliation{Department of Physics and Astronomy, Purdue University, West Lafayette, IN 47907, USA}
\affiliation{Quantum Science Center, Oak Ridge National Laboratory, Oak Ridge, TN 37831, USA}

\author{Kaelyn Ferris}

\affiliation{IBM Quantum, IBM T.J. Watson Research Center, Yorktown Heights, New York 10598, USA}

\author{Xiao Xiao}

\affiliation{North Eastern University, Boston, MA 02115, USA}
\affiliation{Quantum Science Center, Oak Ridge National Laboratory, Oak Ridge, TN 37831, USA}

\author{Lukasz Cincio}
\affiliation{Theoretical Division, Los Alamos National Laboratory, Los Alamos, NM 87545, USA}

\affiliation{Quantum Science Center, Oak Ridge National Laboratory, Oak Ridge, TN 37831, USA}

\author{Nicholas Bronn}

\affiliation{IBM Quantum, IBM T.J. Watson Research Center, Yorktown Heights, New York 10598, USA}

\author{Arnab Banerjee}
\email{arnabb@purdue.edu}
\affiliation{Department of Physics and Astronomy, Purdue University, West Lafayette, IN 47907, USA}
\affiliation{Quantum Science Center, Oak Ridge National Laboratory, Oak Ridge, TN 37831, USA}

\begin{abstract}
Qutrit-based quantum circuits could help reduce the overall circuit depths, and hence the effect of noise, when the system of interest has a local dimension of three. Accessing second excited states in superconducting transmons provides a straightforward hardware realization of qutrits useful for such ternary encoding. In this work, we successfully calibrate microwave pulse gates to a low error rate to operate transmon qutrits. We use these qutrits to simulate one-dimensional spin-1 AKLT states (Affleck, Kennedy, Lieb, and Tasaki), which exhibit a multitude of interesting phenomena, such as topologically protected ground states, string order, and the existence of a robust Berry phase. We demonstrate the efficacy of qutrit-based simulation by preparing high-fidelity ground states of the AKLT Hamiltonian with open boundaries for various chain lengths. We then use ground state preparations of the perturbed AKLT Hamiltonian with periodic boundaries to calculate the Berry phase and illustrate non-trivial ground state topology. To establish the advantage of qutrits over qubits in the presence of noise, we present scalable methods for preparing the AKLT state and computing its Berry phase using tensor network simulations. Our work provides a pathway toward more general spin-1 physics simulations using transmon qutrits, with applications in chemistry, magnetism, and topological phases of matter.
\end{abstract}

\maketitle

\section{Introduction}

Noisy Intermediate‑Scale Quantum (NISQ) \cite{nisq} computers are now actively used to simulate physical systems, with growing impact as they approach practical utility for large‑scale problems \cite{utility,utility2,utility3,utility4,utility5}. Most current simulations rely on qubit‑based encoding, but representing higher‑spin systems (e.g.~\cite{Spin1_condensed_matter,spin1_phy_2,spin1_phy_3,spin1_phy_4}) remains inefficient since spins with $s > 1/2$ must be encoded across multiple qubits. This increases circuit depth and the number of qubits beyond the physical spins under study, amplifying susceptibility to noise and reducing overall fidelity.

Qudit ($d$-level) encoding schemes, which offer access to a larger computational space beyond two levels, have demonstrated advantages in various applications such as error correction (\cite{qutrit_encoding,qutrit_enoding_2}), quantum key distribution (\cite{qkd_1,qkd_2}), and requiring fewer entangling gates in execution of certain algorithms like Shor's \cite{algo_1}, Grover's \cite{algo_2}, and the quantum Fourier transformation \cite{algo_3}. Similarly,  qudit encoding schemes are better suited for simulating systems that are inherently based on $d$ levels.  On the theoretical front, studies aimed at developing tools for qutrit (three-level) simulations, such as algebraic relations for circuit compression and optimization to improve circuit fidelities \cite{qutrit_trotter}, have been undertaken.

Experimentally, significant advances have been made in implementing qutrits using superconducting transmons \cite{transmon2,transmon3}, photonic circuits \cite{photonic}, neutral atoms \cite{4kdl-xcwz}, and trapped ions \cite{trapped_ions,trapped_ions_2}. Several recent multi‑qutrit experiments \cite{entangling_qutrits, Siddiqi_RB,Siddiqi_mitigation} have advanced the state of the art in high‑fidelity gates, benchmarking, and error mitigation on custom‑built hardware, alongside related achievements such as high‑fidelity multi‑qubit gates \cite{galda2021implementing} and maximally entangled qutrit state preparation \cite{Alexey_qutrits}.  Another notable direction is explored in  \cite{transmon1}, which investigates information scrambling and error propagation in qutrit circuits. In principle, one could continue accessing energy levels beyond the third level of transmons to encode higher-spin systems. However, in practical transmon devices, the higher excited states are generally more vulnerable to multiple decoherence mechanisms~\cite{Alexey_qutrits,charge_dispersion}, making them impractical for reliably encoding higher-spin systems unless the Josephson-to-charging energy ratio $(E_J/E_C)$ is significantly increased \cite{wang_25}. As a result, we restrict our focus to qutrits in this work.

\begin{figure*}
\centering
\includegraphics[width=1\textwidth]{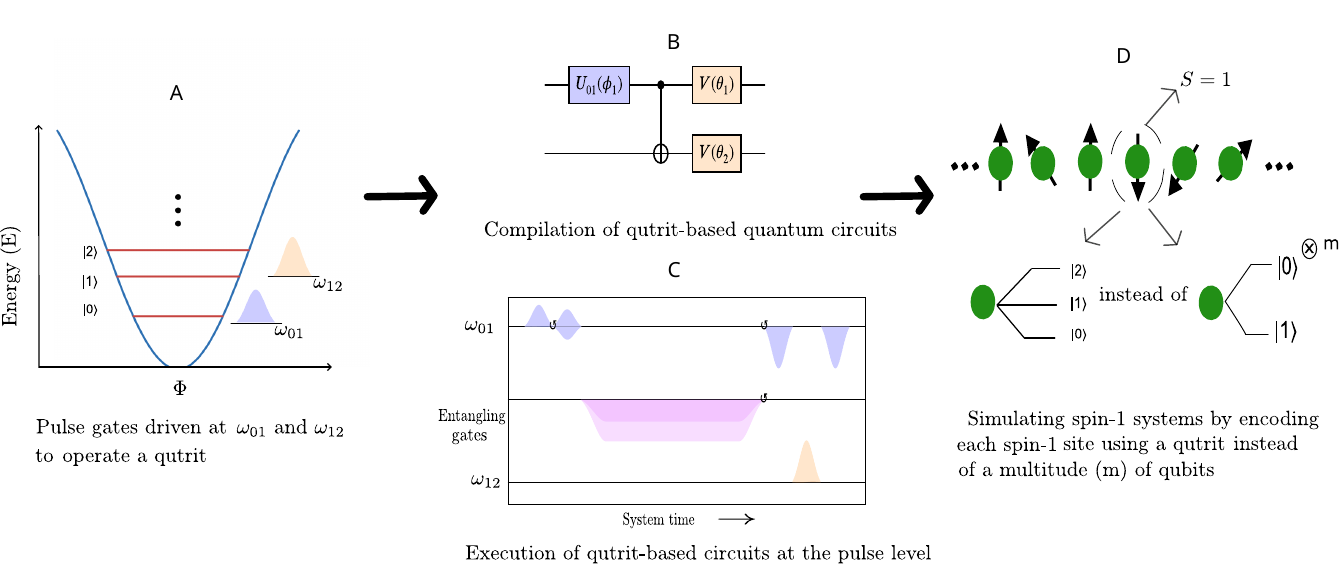}
\caption{The Figure describes a broad overview of this work. (A) We realize qutrits by accessing the second excited states of the transmons by applying calibrated microwave pulse gates driven at $\omega_{12}$ frequency (Section \ref{qutrit gate section}). These pulse gates are used to implement qutrit-based circuits compiled classically (B) using their corresponding pulse schedule (C). (D) We elucidate the pertinency of these transmon qutrit-based circuits by simulating spin-1 systems (AKLT (Section \ref{Theory}) in this work) in Sections \ref{OBC}, \ref{PBC}, and \ref{BP section} which would be costlier in terms of depth and number of entangling gates in the qubit setting. The advantage of using qutrits over qubits is further elucidated by our noisy simulations in a simplified noisy setting.}
\label{fig:Summary}
\end{figure*}

Motivated by the suitability of qutrits for simulating spin‑1 physics, in this work (Fig.~\ref{fig:Summary}) we utilize custom‑calibrated microwave pulses (pulse gates) to access the second excited state of IBM Quantum transmons, enabling the direct simulation of one‑dimensional spin‑1 AKLT \cite{AKLT} systems. 

The one-dimensional spin-1 AKLT Hamiltonian provides one of the simplest symmetry-protected topological models (some others include \cite{2dSPT1,3DSPT1,3DSPT2}) and hosts numerous interesting physical phenomena, such as string order \cite{string_order} and a robust Berry phase \cite{berry_phase_phenom}. There have been numerous investigations undertaken to study these AKLT systems \cite{Sompet_2022,Monroe_1,Monroe_2,optical,PK,DM,qubit1,qubit2}. Recently, a trapped-ion-based simulation of AKLT states was undertaken \cite{edmunds2024constructingspin1haldanephase} aimed at showing the long-range string order of the prepared Open Boundary Condition (OBC) AKLT ground states. While earlier work observed a Berry phase in a single qutrit without an explicit Hamiltonian \cite{martinis_bp}, our study directly probes the perturbed AKLT Hamiltonian, revealing a non‑trivial many‑body Berry phase that provides clear evidence of spin fractionalization into singlets, a feature not previously shown in such a system.

With transmons, earlier ground‑state preparation used qubit‑based encoding and dynamic circuits \cite{qubit1,qubit2}. Expanding the qubit gate set to qutrits, we present a straightforward method for preparing the ground states of the AKLT Hamiltonian. We achieve high‑fidelity preparation with OBC across various chain lengths and with PBC at fixed length under varying perturbations. These perturbations \cite{BP_utility} enable Berry‑phase measurements \cite{BP_utility,BP_utility2,BP_utility3}, offering key insight into valence bonds in the system. Alongside these hardware demonstrations, we present scalable noisy tensor‑network simulations for preparing the AKLT state and computing its Berry phase in spin‑1 chains of up to 40 sites, providing a realistic guide for future experimental implementations. This highlights both the immediate advantages of qutrit‑based circuits over qubit‑based ones in simplified noisy settings and their adaptability to a broader class of spin‑1 systems.

\section{Methods}

\subsection{Qutrit gate set}
\label{qutrit gate section}

Transmons are multilevel systems (with states $\ket{0},\ket{1},\ket{2},...$) typically driven by microwave pulse gates ($X_{01}$,$
\sqrt{X_{01}}$) at the ground state ($\ket{0}$ state) to the first excited state ($\ket{1}$ state) transition frequency. This facilitates encoding of a general normalized two-dimensional complex vector set $\mathbf{B}$ in the $0$–$1$ manifold as follows:

\begin{equation}
\mathbf{B} = \big \{ a_0 \ket{0} + a_1 \ket{1} \;\big|\;  |a_0|^2 + |a_1|^2 = 1 \big \}
\end{equation}

The existence of anharmonicity in the IBM Quantum transmons results in the next transition frequency ($\ket{1}$ to $\ket{2}$) being slightly different compared to the $\ket{0}$ to $\ket{1}$ transition frequency. As a result, applying microwave pulse gates at the $1-2$ frequency ($X_{12}$ and $\sqrt{X_{12}}$) will enable us to encode two-level systems in the $1-2$ manifold. These $1-2$ manifold pulse gates, along with the default $0-1$ manifold counterparts, facilitate encoding a general normalized three-dimensional complex vector set \textbf{T} in the $0-1-2$ manifold as follows:

\begin{equation}
\textbf{T} = \big \{a_0 \ket{0} + a_1 \ket{1} + a_2 \ket{2} \;\big|\   |a_0|^2 + |a_1|^2 + |a_2|^2 =1 \big \}    
\end{equation}
To prepare AKLT ground states, we begin with the default $0$–$1$ manifold gate set, which includes $X_{01}$, $\sqrt{X_{01}}$, the virtual $RZ_{01}$ gate \cite{gambetta}, and the CNOT$_{01}$ gate. When compiling qutrit-based circuits, global phases that were previously insignificant, such as the $-i$ phase associated with the $X_{01}$ gate become relevant. As a result, the full $SU(3)$ matrix representation of these gates must be considered (see Appendix \ref{gates} for the matrix representations of the qutrit gate set used). To extend the gate set, we custom calibrate single-qutrit microwave pulse gates in the $1$–$2$ manifold ($X_{12}$ and $\sqrt{X_{12}}$), and define a virtual $RZ_{12}$ gate. We achieve this using Qiskit Pulse, which allows precise control over the amplitude, frequency, and shape parameters of the microwave pulses \cite{qiskit_pulse,qutrit_cal}.

We follow a detailed calibration procedure to obtain high-fidelity gates in the $1-2$ manifold (see Appendix \ref{calib_details}). After obtaining the reported anharmonicity from the IBM Quantum backend configuration, we refine the $1-2$ transition frequency by performing narrow-band spectroscopy around the estimated value. This is followed by continuous Rabi and Ramsey experiments to obtain the pulse amplitudes of the $X_{12}$ and $\sqrt{X_{12}}$ gates and $1-2$ transition frequency respectively (pulse durations are set to those of the $X_{01}$ gates). These are further refined by error-amplified Rabi and Ramsey experiments~\cite{Sheldon2016} before a final DRAG (Derivative Removal by Adiabatic Gate) calibration~\cite{Motzoi2009} to correct line dispersion and $Z_{12}$ errors from the higher levels of the qutrit~\cite{Jurcevic2021}.  The fidelity of single-qutrit gates in the $1-2$ manifold is then determined by randomized benchmarking~\cite{Magesan2011} in that manifold and reported in Appendix Table \ref{tab:gate_fids}. For completeness, we mention that \cite{Siddiqi_RB} has dealt with a more general randomized qutrit gate benchmarking.

\subsection{Review of the AKLT model}\label{Theory}

The spin-1 AKLT Hamiltonian on a one-dimensional chain of length $n$ is defined as \cite{AKLT}
\begin{equation}
    H = \sum_{i=1}^{n-1} \vec{S}_i \cdot  \vec{S}_{i+1} + \frac{1}{3}(\vec{S}_{i} \cdot \vec{S}_{i+1})^2,
\end{equation}
where the spin-one operator is denoted by the vector operator $\vec{S}_i = (S^x_i,S^y_i,S^z_i)$ with $S^\alpha_i$ representing the spin-1 operator for the $\alpha$-component on the site $i$.  
\begin{figure}

    \centering
    \includegraphics[width=0.8\linewidth]{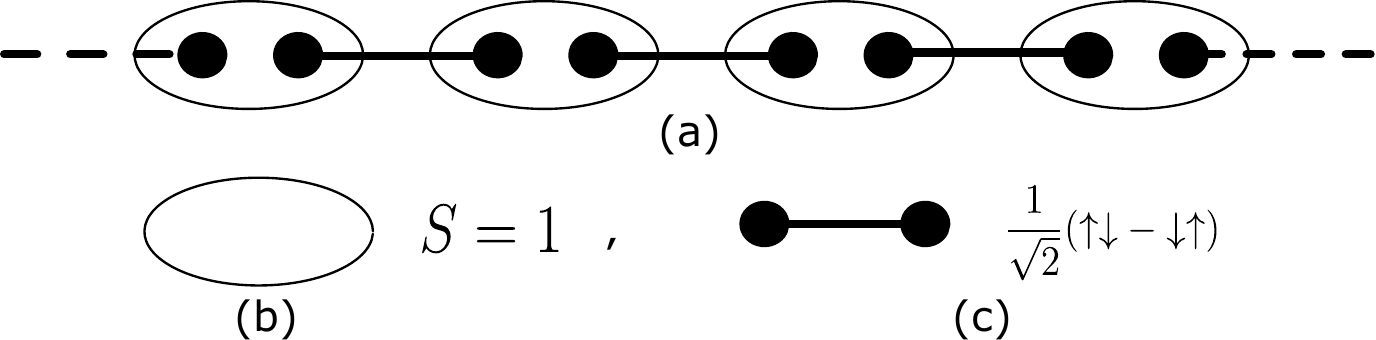}
    \caption{ In spin-1 systems, each spin-1 site can be represented by a tensor product of a pair of spin-1/2 sites (a $\&$ b) followed by their projection to the spin-1 sector.  In AKLT ground states, these adjacent spin-1 sites decompose as pairs of two spin-1/2 sites (a $\&$ c) that form singlets (valence bond states) between them.}
     \label{fig:desc}
\end{figure}

The Hilbert space of the chain $(\mathbb{C}^3)^{\otimes n}$ can be represented by a tensor product of $n$ symmetric subspaces composed of pairs of spin-1/2 sites.  In this representation, the AKLT Hamiltonian is (up to an additive constant) a positive sum of $S=2 $ projectors acting on neighboring site pairs. In the ground states, the spin-1 sites decompose as pairs of spin-1/2 sites that form singlets (valence bond states) between neighboring sites (See Fig. \ref{fig:desc}) have zero $S=2$ projections and hence minimize the Hamiltonian. For OBC, the system has four degenerate ground states corresponding to the Hilbert space of two emergent free spin-1/2 sites at the edges. For PBC, the ground state is non-degenerate as there are no free edges.

The AKLT ground states (both OBC and PBC) also have straightforward matrix product state (MPS) representations. The OBC ground state in this representation has the form

\begin{equation}
    \ket{\Psi}_{OBC} = \sum _{\mathbf{\sigma}} ({b^{l}_{A}}^{T} A^{[\sigma_1]}A^{[\sigma_2]}..A^{[\sigma_N]}b^{r}_{A} ) \ket{\sigma_1\sigma_2..\sigma_n}
    \label{obcgs}
\end{equation}

and for the PBC ground state, we have,

\begin{equation}
    \ket{\Psi}_{PBC} = \sum _{\mathbf{\sigma}} \text{Tr}[A^{[\sigma_1}A^{[\sigma_2]}..A^{[\sigma_N]}]\ket{\sigma_1\sigma_2..\sigma_n} 
    \label{pbcgs}
\end{equation}

where $\sigma_i \in \{0,1,2\}$ labels the $i$-th spin-1 basis state, with corresponding MPS matrices $A^\sigma$ given by
\begin{equation}
    A^{[0]} = -\sqrt{2/3} \tau^{-},  A^{[1]} = -\sqrt{1/3}\tau^z , A^{[2]} = \sqrt{2/3} \tau^{+}
\end{equation}
where $\tau^{\pm} = \tau^x \pm i\tau^y$ and $\tau^z$ span the set of spin-1/2 Pauli matrices. The two boundary vectors $b^r $ and $b^l$ in the Eq \ref{obcgs}  span the four-dimensional degenerate ground state subspace of the OBC AKLT Hamiltonian. We prepare the OBC ground states corresponding to $b^r \otimes  b^l \in 
\{00,01,10,11\}$ in Section \ref{OBC}. These MPS representations play a crucial role in compiling scalable qutrit based ground-state circuits as we will discuss in the Section \ref{scalable_ansatz}. 

\section{Results}
\label{results}
\subsection{OBC AKLT ground states}
\label{OBC}

\begin{figure*}
    \centering
    \includegraphics[width=0.8\textwidth]{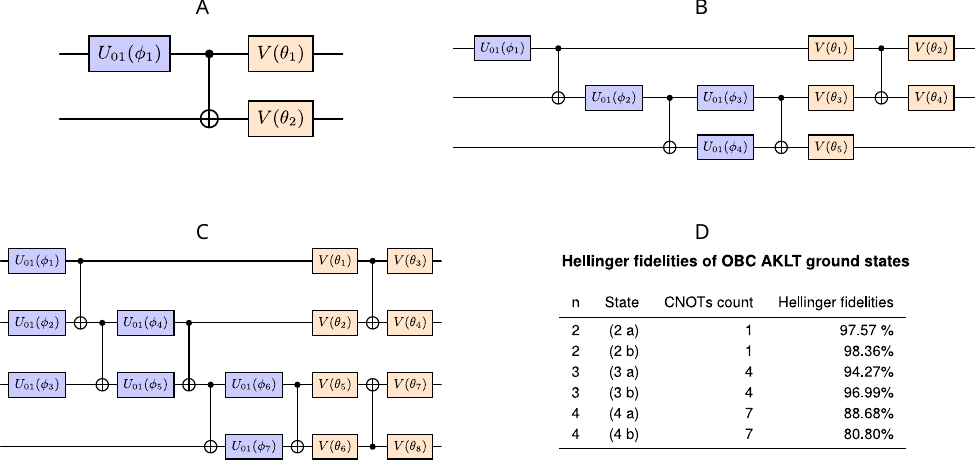}
\caption{\textbf{Circuit ansatzes for OBC AKLT ground states and their hardware Hellinger fidelities:} (A-C) capture all four ground states of $n=2,3$ and $4$ OBC AKLT.  The exact parameters and sequence of pulse gates used in these single-site gates can be found in the Appendix \ref{github} (D) Table contains the Hellinger fidelities (See S2) of all four ground states corresponding to $n=2,3$, and $4$ except for the one ground state of $n=2$, which is fully confined to the $0-1$ manifold and has $99\% - 100 \%$ Hellinger fidelity.} 
\label{fig:obc_hardware}
\end{figure*}

The circuit ansatzes for chain lengths $n=2,3,4$  are depicted in Fig. \ref{fig:obc_hardware}  A, B, and C. These circuits are designed to prepare the degenerate ground states of the OBC AKLT chain using qutrit-based gates. However, working with higher-dimensional states would bring in additional channels of decoherence such as $\ket{2} \rightarrow \ket{1}$ relaxation ($T_1$ process), dephasing ($T_2$), and aggravation of qubit crosstalk due to the inclusion of $\ket{1} \rightarrow \ket{2}$ transition frequencies \cite{Alexey_qutrits}. This leads us to consider circuits where we restrict all the possible $0-1$ manifold operations ($U_{01}(\phi)$ in Fig. \ref{fig:obc_hardware} A, B and C) at the start of the circuit and reserve the qutrit pulse gates/operations, $V(\theta)$ (Refer to Appendix \ref{github} for the precise sequences of pulse gates/operations)  toward the end of the circuits. Such circuit ansatzes capture all four degenerate ground states of an OBC AKLT chain for chain lengths up to four while making sure that the effects of charge dispersion \cite{charge_dispersion} and ac-Stark shifts due to higher levels are minimized. If high-fidelity entangling qutrit gates are available (as in, e.g., \cite{entangling_qutrits}), then these additional considerations become unnecessary, as the scalable ground-state ansatz described in Sections \ref{scalable_ansatz} is sufficient to maximize the fidelities.

We obtain the Hellinger fidelities (see Appendix \ref{definitions}) of the OBC AKLT ground states with the scheme in Appendix \ref{additional_circs} and list them in the table in Fig. \ref{fig:obc_hardware} D . For $n=2$, one of the ground states is confined to $0-1$ manifold and can be prepared without qutrit gates to  $\sim 99\% - 100 \%$ Hellinger fidelity.  For the other three ground states, circuits (2 a) and (2 b), along with $X_{01}$ gates appended circuits (as described in Appendix \ref{additional_circs}) suffice for Hellinger fidelity computation. Similarly,  circuits (3 a) and (3 b) and circuits (4 a) and (4 b), along with the $X_{01}$ gates appended circuits suffice to compute the Hellinger fidelities of all the four ground states corresponding to $n=3$ and $n=4$ respectively. Thus, we report Hellinger fidelities for two representative states per chain length.

We observe that with the increase in chain lengths, the circuits become deeper and noisier, resulting in lower fidelities. Error mitigation techniques developed in \cite{Siddiqi_mitigation} may help extend our state preparation to deeper circuits, even though their applicability to IBM’s hardware remains to be tested. We also expect improvements from approaches such as dynamical decoupling aimed at reducing the $\ket{2} \rightarrow \ket{1}$ decay, as explored in \cite{Alexey_qutrits,galda2021implementing}.

\subsection{Perturbed PBC AKLT ground states}
\label{PBC}
The Hamiltonian for a 1D chain of length $n$ with periodic boundary conditions is expressed as follows
\begin{equation}
    \begin{split}
   H = \sum_{i=1}^{n-1} & \left( \vec{S}_i \cdot \vec{S}_{i+1} + \frac{1}{3}(\vec{S}_i \cdot \vec{S}_{i+1})^2 \right) \\
    & +\left( \vec{S}_n \cdot \vec{S}_{1} + \frac{1}{3}(\vec{S}_n \cdot \vec{S}_{1})^2 \right) \ .
   \end{split}
\label{PBC_eq}   
\end{equation}

In the PBC system represented by Eq.~\ref{PBC_eq}, a class of Hamiltonians can be defined by introducing local perturbations to the bond between the $j$-th and ($j+1$)-th sites.  We modify the Hamiltonian for the perturbed bond as follows,

\begin{equation} \label{perturbed bond}
    \vec{S}_j \cdot \vec{S}_{j+1} \rightarrow
    e^{i\theta} S_{j}^{+}S_{j+1}^{-}/2 + e^{-i\theta} S_{j}^{-}S_{j+1}^{+}/2+ S^{z}_{j}S^{z}_{j+1} \ ,
\end{equation}
where $S_i^{\pm} = S_i^x \pm S_i^y$. This perturbation, proposed in~\cite{BP_utility}, is a local spin twist and leaves other bonds unchanged.
This twist also introduces a generic parameter essential for computing the Berry phase in AKLT-like systems.

\begin{figure*}[h!]
  \centering
  \includegraphics[width=1\textwidth]{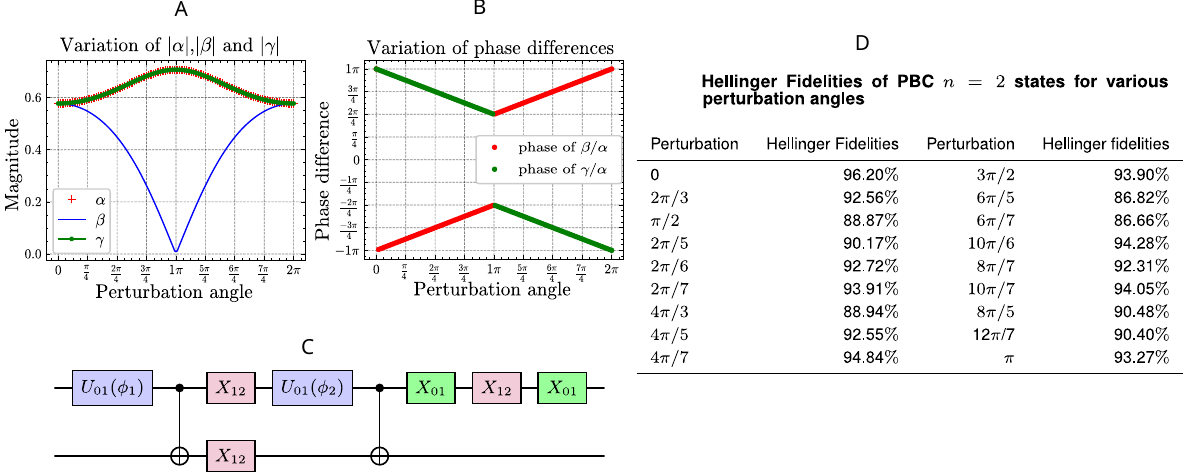}
  \caption{\textbf{Ground states of Perturbed PBC AKLT ground states}  The ground state vector $\alpha(\theta)$$\ket{02}$ + $\beta(\theta)$$\ket{11}$
  + $\gamma(\theta)$ $\ket{20}$'s dependence on the perturbing angle $\theta$ is described in Figures A and B. We note that $|\alpha| = |\gamma|$ for all perturbation angles. At $\theta = \pi$, we observe that $\beta = 0$, coinciding with the point where the phase difference experiences a sudden discontinuity. (C) $\mathbf{n=2}$ \textbf{PBC ansatz} captures the perturbed PBC ground states of $n=2$ AKLT for all perturbations. However,  for $\theta = \pi$, ansatz with just a single entangling gate exists (See Fig. \ref{fig:obc_hardware} A). (D) Table contains the Hellinger fidelities of the  $n=2$ perturbed PBC AKLT states for various perturbation angles. These angles are used for the computation of the Berry Phase in Section \ref{berry_hardware}.}
  \label{PBC_hardware}
\end{figure*}

Any finite-length PBC AKLT chain exhibits a $\pi$ Berry phase \cite{AKLT-statesZX} under the perturbation described in Eq.~(\ref{perturbed bond}). To demonstrate this on hardware, we prepare the perturbed ground states on chains of length 2 using an ancilla-free ansatz (Fig. \ref{PBC_hardware} C).  Unlike the OBC case, our focus here is limited to short chains due to the absence of closed-loop connectivity in current hardware, which restricts the implementation of longer periodic chains without ancillary qubits. To address this limitation, we explore noisy simulations using a scalable ansatz containing ancillary qubits in Section \ref{scalable_ansatz}, which circumvents the need for closed-loop connectivity and enables extension to longer chains.

For $n=2$, the ground state of the perturbed AKLT Hamiltonian under PBC can be expressed as follows: 
\begin{equation}
    \ket{GS(\theta)} = \alpha(\theta) \ket{02} + \beta({\theta}) \ket{11} + \gamma(\theta) \ket{20} 
    \label{PBC_state}
\end{equation}

Without perturbation ($\theta = 0$), the ground state vector is $\frac{1}{\sqrt{3}}\ket{02} - \frac{1}{\sqrt{3}}\ket{11} + \frac{1}{\sqrt{3}}\ket{20}$ is maximally entangled. At $\theta = \pi$, the ground state  vector is $\frac{1}{\sqrt{2}}\ket{02}-\frac{1}{\sqrt{2}}\ket{20}$. This state can be prepared with high fidelity using the ansatz for the OBC state in Fig.~\ref{fig:obc_hardware} A. All other perturbations possess non-zero support on all the three components $\ket{02},\ket{11}$, and $\ket{20}$. These states can be prepared to a high Hellinger fidelity (see the table in Fig~\ref{PBC_hardware} D) using the ansatz circuit shown in Fig.~\ref{PBC_hardware} C.

 While the ansatz Fig.~\ref{PBC_hardware} C can be used to prepare a high Hellinger fidelity state, it is not guaranteed that the relative phase differences between various components are captured accurately. This is because, in the second entangling gate of the ansatz, the control state is in general ($0-1-2$ ) manifold and this results in a non-negligible Stark phase shift amongst the statevector components. Accurately capturing these phase differences is essential for correctly estimating the Berry phase. We address this issue through state tomography and phase correction techniques, as detailed in Appendix \ref{phase_experiments}.

\subsubsection{Scalable state preparation}
\label{scalable_ansatz}

\begin{figure*}
\centering
    \includegraphics[width=0.95\textwidth]{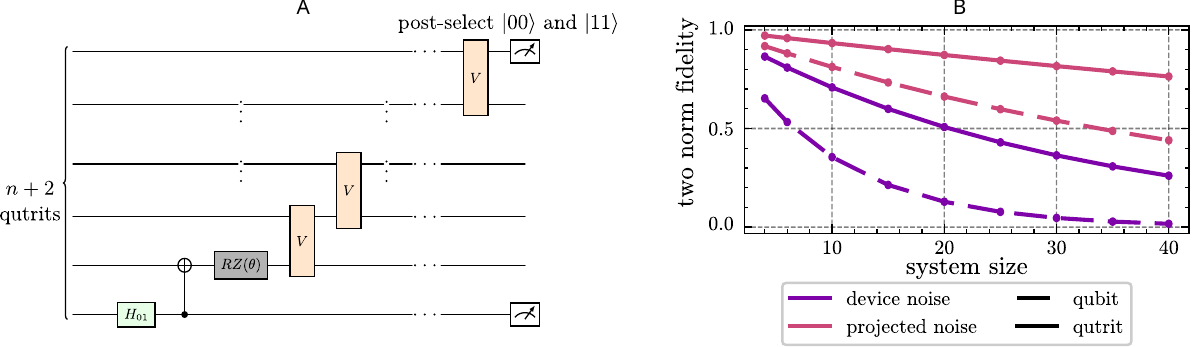}
    \caption{\textbf{(A) Scalable AKLT state preparation}. The ground state of the PBC AKLT model with perturbation angle $\theta$ is generated from an equal superposition of the $\ket{00}$ and $\ket{11}$ post-selected states. Note that for qubit-based implementation, each bulk qutrit must be encoded into two qubits. The $H$ and $RZ$ gates are standard qubit gates on a qubit device and the equivalent 0-1 manifold qutrit gates on a qutrit device. The entangling operation $V$ can be derived from the tensors of the MPS representation of the ground state. \textbf{(B) Simulation of state preparation in the presence of noise}  Qutrit-based implementation consistently has higher fidelity than qubit-based implementation. The projected noise model uses $T_1$ and $T_2$ times that are five times longer than the ones measured on the physical device.}
    \label{scalable_fidelity}
\end{figure*}
The AKLT ground states can be represented as matrix product states (MPS) of bond dimension two. This motivates the linear-depth circuit ansatz \cite{mps_encoding} described in Fig.~\ref{scalable_fidelity} A, where the repeated entangling operation $V$ can be derived from the tensors of the MPS (see Eq. \ref{pbcgs}). This operation can then be compiled onto the gate set available on either qutrit- or qubit-based hardware. Note that for the qubit-based compilation, we encode each bulk qutrit into a pair of qubits and use the following mapping of computational basis states: $\{\ket{0}\to\ket{00}, \ket{1}\to\ket{01}, \ket{2}\to\ket{10}\}$. More details regarding the compilation procedure can be found in Appendix \ref{circuit_compilation}. 

The circuit described in Fig.~\ref{scalable_fidelity} A prepares the AKLT ground state by measuring the first and last qudits and post-selecting the outcomes:
\begin{align}
    \ket{GS(\theta)} &= \frac{1}{\mathcal{N}_{00}} \langle 00|\psi(\theta)\rangle + \frac{1}{\mathcal{N}_{11}} \langle 11|\psi(\theta)\rangle \ ,
\end{align}
where $\ket{\psi(\theta)}$ is the pre-measurement state, and $\mathcal{N}_{00}, \mathcal{N}_{11}$ ensure normalization. We numerically determine that the probabilities of outcomes $00$ and $11$ are approximately 0.25, irrespective of system size or perturbation angle. Therefore, this method for state preparation has only a small sampling overhead. This linear-depth ansatz could be improved to constant depth circuits with the help of mid-circuit measurements and feed-forward operations at the cost of a constant factor in qudit count as noted in \cite{qubit2,const_depth}.

To illustrate the utility of qutrits, we simulate implementations of this state preparation method on qutrit- and qubit-based hardware in the presence of realistic noise. Our noise models (see Appendix \ref{noise_model}) consider the thermalization of qubits and qutrits via the amplitude damping and dephasing channels, and can be simulated efficiently using tensor network techniques. We choose the two-norm fidelity (Appendix \ref{definitions}) as the metric of success for state preparation. 

Figure~\ref{scalable_fidelity} B shows that as we scale up system size, qutrits lead to preparations with significantly higher fidelity, both for noise rates available on existing devices and noise rates projected to be available on devices in the near future. This is because the compilation of the entangling operation, $V$, requires a larger number of CNOT gates on a qubit device.  

\begin{figure*}[t]
  \centering
  \includegraphics[width=1\textwidth]{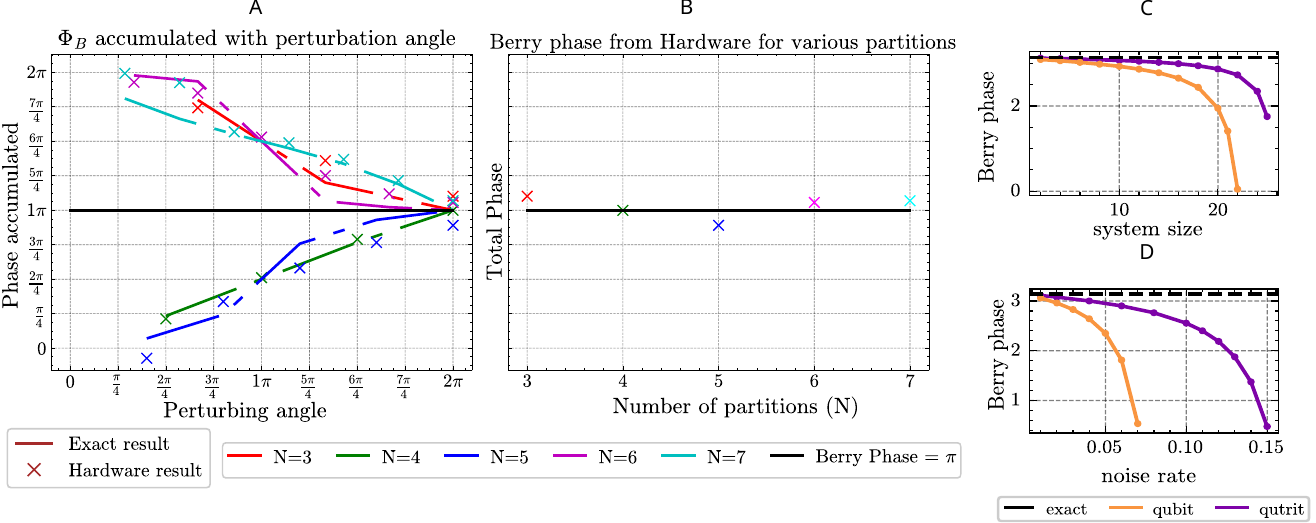}
  \caption{\textbf{Berry phase from quantum hardware:} Panel (A) shows the accumulated phase when the ground state is perturbed as described in the main text (from perturbation angle 0 to $2\pi$) for various numbers of partitions. Common, randomly chosen gauge vectors were selected so that the partial sums in Eq.~\ref{discrete berry} obtained from the quantum hardware (plotted with $\times$ markers) could be directly compared with the noiseless classical results (plotted as dashed lines). Panel (B) represents the final values of Berry phase for various numbers of partitions. \textbf{Nosiy Simulation of Berry phase computation} (C) across system sizes and (D) across noise rates. The Berry phase remains close to the expected value of $\pi$ for a range of system sizes and noise rates before sharply decaying to zero. We identify this as a phase transition in the learnability of the topological properties of the AKLT state. For panel (C), the noise rate was fixed at 0.01, and for panel (D), the system consisted of 4 spins. All the Berry phases in (C) and (D) were computed using 4 partitions of the [0,2$\pi$] interval of perturbing angles.}
  \label{bp_combined}
\end{figure*}

\subsection{Berry phase}
\label{BP section}
\label{BP_comp}
The topology of ground states of a spin-1 chain with time-reversal symmetry can be characterized by introducing a Berry phase for locally perturbed bonds \cite{Hatsugai_2007,BP_utility}. This differs from other methods for characterizing topological phases \cite{Pollmann_2012}, which require extraction of the density matrix for a subsystem much larger than the correlation length of the system.  Since Berry phase estimation is based on introducing a local perturbation, it is easier to implement on NISQ hardware.  

For a parameter-dependent Hamiltonian, $H(\theta)$, the Berry phase $\gamma$ of the non-degenerate ground state is defined as 
\begin{equation}
    \label{Berry integral}
    \gamma = -i \int _{0} ^{2\pi} A(\theta) d\theta,
\end{equation}
where $A(\theta)$ is the Abelian Berry connection obtained from the normalized ground state $\ket{GS(\theta)}$ of $H(\theta)$ as $A(\theta) = \bra{GS(\theta)}\partial_{\theta} \ket{GS(\theta)}$ such that $\ket{GS(0)} = \ket{GS(2\pi)}$. At zero perturbation, we introduce a slight modification to the coefficient of the quartic term in Eq. \ref{PBC_eq} to break the degeneracy of the ground states and eliminate ambiguity in the computation of Eq. \ref{Berry integral}. Specifically, we adjust the coefficient from $\frac{1}{3}$ to $\frac{1}{3} - \epsilon$, where $\epsilon$ is a small positive parameter. This ensures that we focus exclusively on non-degenerate ground states. For a detailed treatment of degenerate systems, see \cite{Hatsugai_multiplets}.

We note that the Berry phase is real and quantized to $0$ or $\pi$ (mod $2\pi$) if the Hamiltonian $H(\theta)$ is invariant under any anti-unitary operation $\Theta$, \textit{i.e.} $[H(\theta),\Theta] = 0$. The perturbation defined in Eq. \ref{perturbed bond} has time-reversal symmetry and, therefore, satisfies this condition. A Berry phase of $\pi$ resulting from a perturbation on a bond between neighboring sites indicates the presence of a valence bond state, which has a non-trivial topology and entanglement, which is the central feature we aim to capture.

To calculate the Berry phase numerically, we motivate a discrete approximation of it by partitioning the integration region in Eq. \ref{Berry integral} with $N-1$ points between $0$ to $2\pi$ 
\begin{eqnarray*}
    \bra{GS(\theta)} \partial_{\theta} \ket{GS(\theta)} \simeq \ln[1 + \delta \theta \bra{GS(\theta)}\partial_{\theta} \ket{GS(\theta)}] \\
    \simeq \ln \langle {GS(\theta)}|{GS(\theta + d\theta)}\rangle  = -i \;\text{Arg}\langle {GS(\theta)} |{GS(\theta + d\theta)} \rangle
\end{eqnarray*}

With this approximation, we can write a discretized expression for the Berry phase
\begin{equation}
    \label{discrete berry}
    \gamma_N = - \sum _{j=0}^{N-1} \;\text{Arg} \langle {GS(\theta_j)} |{GS(\theta_{j+1})} \rangle, \theta_j = \frac{2\pi}{N}j
\end{equation}

To compare the accuracy of the partial sums in the Berry phase computation with classical methods such as exact diagonalization or density matrix renormalization group, it is essential to compute the phases using the same gauge for all methods. This is called gauge fixing and is done by picking a random gauge vector $\ket{g}$ and replacing ${|GS(\theta)\rangle}$ with $\langle{GS(\theta)|}{g}\rangle{|GS(\theta)}\rangle$. This procedure fixes any global phase difference between the states produced by classical and quantum methods and allows us to compare their partial sums and not just the (previously) gauge invariant $\gamma_N$. Refer to the Github repository linked in Section \ref{github}.

\subsubsection{Berry phase from quantum hardware}
\label{berry_hardware}

Statevectors corresponding to the perturbing angles shown in the table in Fig.~\ref{PBC_hardware} D are sufficient to compute Eq.~\ref{discrete berry} using 3 to 7 partitions of the $[0, 2\pi]$ interval. The circuit in Fig.~\ref{PBC_hardware} C can be used to prepare, with high Hellinger fidelity, these states. Relative phase differences were applied exactly between the dominant (in magnitude) components, as these can be measured using state tomography and corrected with high accuracy involving only the addition of virtual, noiseless $RZ$ gates (see Appendix \ref{phase_experiments}, specifically Fig.~\ref{phase_rot} and Fig.~\ref{phase_corr}). The phases attached to non-dominant components are noisy, and due to their small magnitudes, their contribution to the overall Berry phase is negligible. As a result, we bypass experiments designed to capture the phase differences among the non-dominant terms and disregard their contribution to the overall Berry phase. This approach can be seen as post-selecting only the dominant components and calculating the total Berry phase using them.

Fig. \ref{bp_combined} A presents the results of our experiment. The resulting Berry phases agree with the expected value of \(\pi\), with discrepancies of approximately \(10\%\) across different choices of partitions. The observed accuracy in the Berry phase and the partial sums (after gauge fixing) can be attributed to the high fidelity of state preparation (table in Fig. \ref{PBC_hardware} D).

 The convergence of the Berry phase to $\pi$ for more than three partitions can be confirmed through classical simulation. To compute the Berry phase of general spin-1 systems, one must first run an analysis on the required number of partitions to observe convergence to the expected value. Computing individual phase differences (and eventually the Berry phase) for larger system sizes using this method will be expensive, although classical shadows \cite{shadows} provide a possible means of reducing the number of circuits required. However, a more straightforward approach is to use a modified Hadamard test alongside the state preparation scheme we described in Section \ref{scalable_ansatz}. We explore this possibility next. 

\subsubsection{Scalable computation of Berry phase}
\label{noisy_bp}

The Hadamard test is ideal for computing $\text{Arg} \langle {GS(\theta_j)} |{GS(\theta_{j+1})} \rangle$ in Eq.~\ref{discrete berry}. It requires input circuits that prepare $\ket{GS(\theta_j)}$. We propose using the circuit in Fig.~\ref{scalable_fidelity} A. However, the post-selection step therein is incompatible with the standard Hadamard test since it requires performing a conditional measurement and reset, which is not a typical operation in the standard Hadamard test. Therefore, we construct a modified Hadamard test that can work with post-selection based state preparation schemes. This algorithm is described in detail in Appendix \ref{gen_hadamard_test}. 

Panels C and D of Figure~\ref{bp_combined} show a simulation of our algorithm being implemented on qutrit- and qubit-based devices in the presence of noise. We note that for a range of system sizes and noise rates (defined as the ratio of coherence times measured on the device and coherence times used for simulation), the qutrit-based implementation performs better, further confirming the advantage of using qutrits to study spin-1 systems. 

We also note that, for both qubits and qutrits, the Berry phase remains robust for a range of parameters before abruptly decaying to zero. We interpret this as a phase transition in the learnability of topological properties as a function of noise. This observed phenomenon might be of independent scientific interest as a new setting for studying learnability transitions \cite{ippoliti_transition}. 

\section{Conclusion}
We demonstrated that by deliberately coming up with circuits that minimize the noise due to higher levels, one can tap into the full potential of qutrit-based encoding schemes. This was demonstrated through high-fidelity state ground state preparations and Berry phase estimation from hardware. We then extrapolated the utility of qutrit-encoding schemes for larger systems through noisy tensor network simulations. 

The methods discussed here could be applicable beyond the specific spin-1 systems, such as spin-1 physics simulations in high energy physics \cite{App_hep_1,App_hep_2,App_hep_3},  condensed matter physics \cite{App_stat_1,App_stat_2,rydberg, fermi_hubbard}, and cosmology \cite{App_cosmo_1}. Another interesting direction is the study of time evolution of spin-1 systems. Qutrit-encoding of such problems requires a modification to conventional Trotter approaches \cite{qutrit_trotter}. The methods described in this work could open up new avenues for the study of the dynamics of spin-1 systems. This constitutes a potential future step in our work.

\section{Acknowledgments}

We would like to acknowledge the funding from the Office of Science through the Quantum Science Center (QSC), a National Quantum Information Science Research Center. We acknowledge the use of IBM Quantum services for this work. L.C. was partially supported by the Laboratory Directed Research and Development program of Los Alamos National Laboratory under project number 20230049DR. A.B. acknowledges Paul Kairys for insightful discussions related to this work. This research used resources of the Oak Ridge Leadership Computing Facility, which is a DOE Office of Science User Facility supported under Contract DE-AC05-00OR22725.\\

\section{Code Availability}

Details regarding the exact states, sequence of pulse gates applied, and the hardware data corresponding to the results shown in the text can be found in the GitHub repository linked in Appendix \ref{github}.

\bibliography{ref}

\widetext
\pagebreak

\appendix

\AtAppendix{\counterwithin{lemma}{section}}

\section{Matrix representations of qutrit gates}
\label{gates}
Matrix representations of single qutrit gates used in the compilation of ground-state circuits.\\
\begin{equation}
        X_{01} = \begin{bmatrix}
0 && -i && 0 &\\\\
-i && 0 && 0 &\\\\
0 && 0 &&  1 &
\end{bmatrix} 
,\hspace{70pt}
      X_{12} = \begin{bmatrix}
1 && 0 && 0&\\\\
0 && 0 && -i&\\\\
0 && -i && 0
\end{bmatrix} \\
\end{equation}
\begin{equation}
  \sqrt{X_{01}} = \begin{bmatrix}
\frac{1}{\sqrt{2}} && \frac{-i}{\sqrt{2}} && 0 &\\\\
\frac{-i}{\sqrt{2}} && \frac{1}{\sqrt{2}} && 0 &\\\\
0 && 0 && 1
\end{bmatrix} 
,\hspace{50pt}
    \sqrt{X_{12}} = \begin{bmatrix}
  1 && 0 && 0 &\\\\
0 && \frac{1}{\sqrt{2}} && \frac{-i}{\sqrt{2}} &\\\\
0 && \frac{-i}{\sqrt{2}} && \frac{1}{\sqrt{2}} &\\\\
\end{bmatrix}
\end{equation}
\begin{equation}
   RZ_{01}(\phi) = \begin{bmatrix}
1 && 0 && 0&\\\\
0 && e^{(i\phi)} && 0&\\\\
0 && 0 && 1
\end{bmatrix} 
,\hspace{40pt}
    RZ_{12}(\phi) = \begin{bmatrix}
1 &&0 && 0&\\\\
0 && 1 && 0&\\\\
0 && 0 && e^{(i\phi)}
\end{bmatrix} 
\end{equation}

The truth table of CNOT gate used in the ground state preparation circuits compilation (Followed from \cite{Alexey_qutrits})

\begin{table}[h]
    \caption{Truth table of default IBM CNOT gate}
    \centering
    \begin{tabular}{|c|c|c|}
      \hline
        Control & Target & Output\\
        \hline
        $\ket{0}$ & $\ket{0}$  & $\ket{00}$\\
                 $\ket{0}$ & $\ket{1}$  & $\ket{01}$\\
                 $\ket{0}$ & $\ket{2}$  & $\ket{02}$\\
                 $\ket{1}$ & $\ket{0}$  & $\ket{11}$\\
                 $\ket{1}$ & $\ket{1}$  & $\ket{10}$\\
                 $\ket{1}$ & $\ket{2}$  & $i\ket{12}$\\
                 $\ket{2}$ & $\ket{0}$  & $a\ket{20} + b\ket{21}$ \\
                 $\ket{2}$ & $\ket{1}$  & $b^*\ket{20} + c\ket{21}$\\
                 $\ket{2}$ & $\ket{2}$  & $e^{i\delta}\ket{22}$\\
        \hline
    \end{tabular}
    
    \label{truth_table}
\end{table}
Some parts of our circuits involve the application of $U$ gate on the qutrits completely in the $0-1$ manifold. We call such gates as $U_{01}$ gate and their matrix representation is as follows\\

\begin{equation}
  U_{01}(\theta,\phi,\lambda) = \begin{bmatrix}
\cos (\theta/2) && - e^{i\lambda} \sin (\theta/2) &\\\\
 e^{i\phi} \sin (\theta/2) && e^{i(\phi+\lambda)} \cos(\theta/2) &\\\\
\end{bmatrix}
\end{equation}
The values of $a$, $b$, $c$, and $\delta$ vary based on the device and the qutrit pairs of interest. For the purpose of noisy simulations performed in this work, we chose $a= 1/\sqrt{2}, b= -1/\sqrt{2}, c=-1/\sqrt{2}$ and $\delta= -\pi /2$. 

\section{Fidelity definitions}
\label{definitions}

\textbf{Hellinger fidelity }of two probability distributions/vectors  $Q_1 = (Q_1(1), Q_1(2),..., Q_1(n))$
and  $Q_2 = (Q_2(1), Q_2(2),..., Q_2(n))$  with n elements each is given by
\begin{equation}
    H = \left(\sum_{i=1}^n \sqrt{Q_1(i)\times Q_2(i)} \right)^2
\end{equation}

\textbf{Quantum fidelity} between states $\rho$ and $\sigma$ is defined as
\begin{equation}
    F(\rho, \sigma) = \left(\text{tr}\sqrt{\sqrt{\rho}\sigma\sqrt{\rho}}\right)^2. 
\end{equation}
Quantum fidelity can be interpreted as a distance measure on probability distributions represented by density matrices. It also generalizes the notion of overlaps to mixed states. However, it is intractable to compute for states of many particles, even when the states have efficient tensor network representations. This is because the square root in the definition requires all the eigenvalues of the exponentially large density matrices.  \\

\textbf{Two-norm fidelity} is defined using the Schatten 2-norm, $\lvert\lvert\cdot\rvert\rvert_2$, on the vector space of operators: 
\begin{align}
    F_2(\rho,\sigma) &= \frac{1}{2} \left(\text{tr}(\rho^2) + \text{tr}(\sigma^2) - ||\rho-\sigma||_2 \right) = \text{tr}\left(\rho\sigma\right)
\end{align}
While two-norm fidelity does not have an interpretation as a statistical distance, it is easy to compute for density matrices represented as matrix product operators of low bond dimension. In fact, it can be expressed as a standard contraction of two matrix product operators. \\

We further justify the use of two-norm fidelity as a metric of success in our numerical simulations by showing that it approximates the quantum fidelity in the low-noise regime. Suppose $\rho$ and $\sigma$ are almost pure states, in the sense that they each have diagonalizations of the form: 
\begin{align}
\rho = (1-\epsilon)\ket{\psi_1}\bra{\psi_1} + \frac{\epsilon}{N-1}\sum_{j=2}^N\ket{\psi_j}\bra{\psi_j} \\
\sigma = (1-\epsilon)\ket{\phi_1}\bra{\phi_1} + \frac{\epsilon}{N-1}\sum_{j=2}^N\ket{\phi_j}\bra{\phi_j},
\end{align}
for small $\epsilon$. Note that the density matrices do not have to commute and the bases in which they are diagonal can be different. The purity of each matrix is approximately $(1-\epsilon)^2$: 
\begin{align}
\text{tr}\left(\rho^2\right) = \text{tr}\left(\sigma^2\right) = (1-\epsilon)^2 + \mathcal{O}\left(\frac{\epsilon^2}{N}\right).
\end{align}
We can now expand quantum fidelity and two-norm fidelity in powers of $\epsilon$: 
\begin{align}
F(\rho,\sigma) &= (1-\epsilon)^2 |\langle  \psi_1 | \phi_1 \rangle|^2 + \mathcal{O}\left(\frac{\sqrt{\epsilon}}{N-1}\right) \\ 
F_2(\rho,\sigma) &= (1-\epsilon)^2 |\langle \psi_1| \phi_1 \rangle|^2 + \mathcal{O}\left(\frac{\epsilon}{N-1}\right)
\end{align}
We observe that the two expressions are equal to order $\sqrt{\epsilon}$. In the low-noise regime, states have purities close to 1; therefore, the two-norm fidelity is a good approximation for quantum fidelity. 

\newpage
\section{Device specifications}
\label{device_specs}
\begin{table}[h!]
\caption{\textbf{Calibration details of \textit{ibm\_hanoi}}}
\centering
\begin{tabular}{||c | c | c | c|c||} 
 \hline
 Qubit $\#$: & Qubit 1 & Qubit 2 & Qubit 3 & Qubit 4 \\ [0.5ex] 
 \hline\hline
 Frequency (GHz) & 5.155 & 5.256 & 5.097&5.0728\\ 
 Anharmonicity(GHz)&-0.3421&-0.3396&-0.3427&-0.3429\\
 $T_1 (\mu s)$ & 146.485 & 173.626 & 140.629& 163.587\\
 $T_2 (\mu s)$ & 145.647& 221.524 & 28.345&14.850\\
 Readout assignment error & 0.0109 & 0.0177 & 0.009&0.0086 \\ [1ex] 
 \hline
\end{tabular}\\
\begin{flushleft}
\caption{Listed in this table are some of the calibration details of \textit{ibm\_hanoi} when using it to prepare high-fidelity AKLT ground states. For $n=2$ (both OBC and PBC (Fig.~\ref{fig:obc_hardware} A and Fig.~\ref{PBC_hardware} C in the main text), qubits 1 and 2 were used. For $n=3$ (Fig.~\ref{fig:obc_hardware} B in the main text), qubits 1, 2 and 3 were used. For $n=4$ (Fig.~\ref{fig:obc_hardware} C in the main text), qubits 1, 2, 3, and 4 were used. Qubits 3, 2, 1, and 4 exist in a single linear chain in that order.}
    
\end{flushleft}
\label{table:qubit_calibs}
\end{table}

\section{Calibration details}
\label{calib_details}

To achieve a high-fidelity $X_{12}$ gate, which is a $\pi$ rotation in the 1-2 manifold about the corresponding x-axis of the Bloch sphere, we start with narrow-band spectroscopy to get a rough estimate of the $1-2$ frequency. This is followed by a more detailed calibration procedure at around this frequency as described below. The following experiments are performed sandwiched between two $X_{01}$ gates so that they occur in the $1-2$ manifold of the qutrit.\\

 \subsection{Rough Rabi experiment}
\begin{figure}[htbp]
  \centering
  \includegraphics[width=0.7\textwidth]{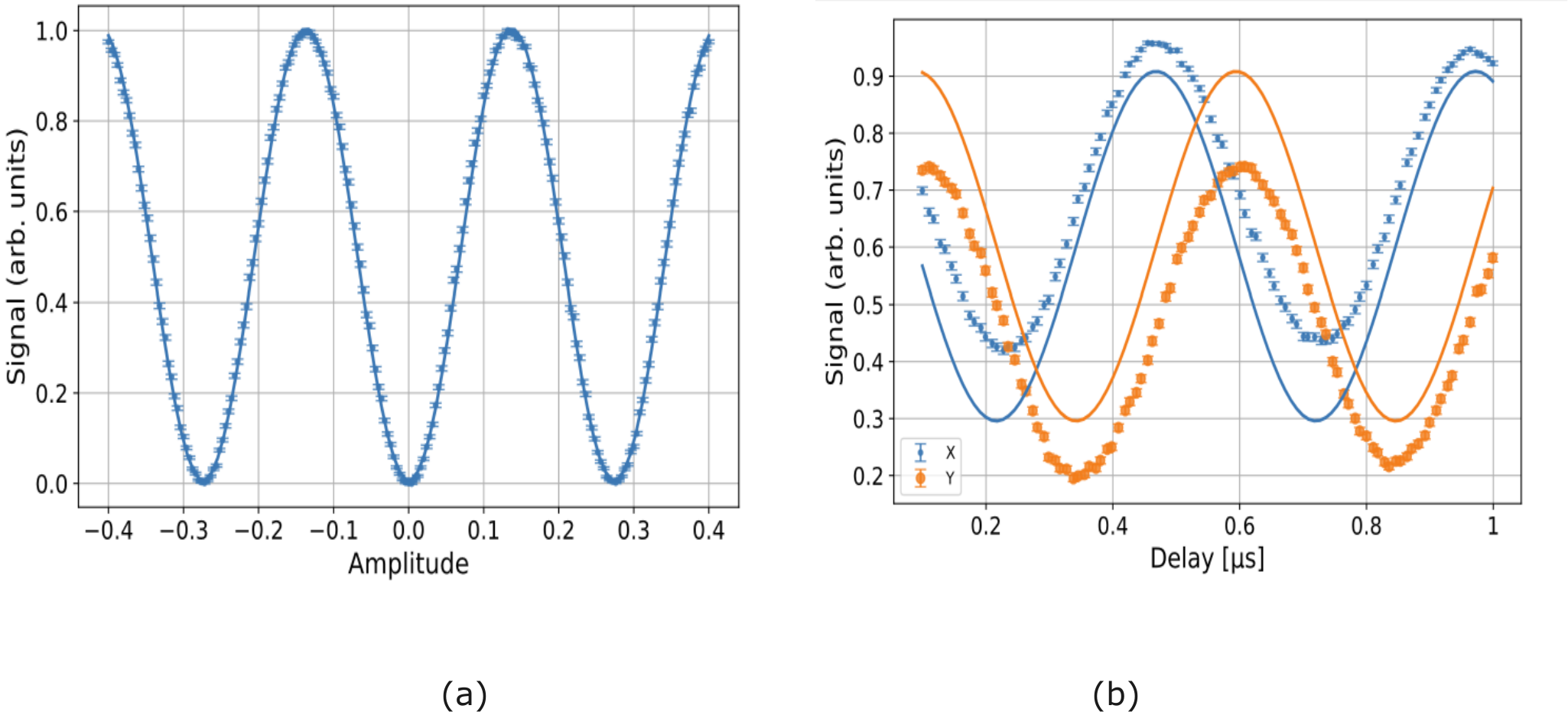}
  \caption{Rough Rabi (a) and Rough Ramsey (b) experiments on the  $1-2$ manifold. The plot here shows a run on the qubit 4  of the \textit{ibm auckland} device. These experiments are followed by fine Rabi and fine Ramsey experiments to fine-tune the amplitude and frequency required for the $X_{12}$ drive pulse. The Ramsey experiments are offset due to the presence of charge dispersion in the $1-2$ transition frequency~\cite{charge_dispersion}.} \label{calibration experiments}
\end{figure}

 A Gaussian pulse of the same duration as a single qubit gate is applied at the estimated $1-2$ frequency by varying the amplitude of the drive, as shown in Fig.~\ref{calibration experiments}(a). The amplitude at which the signal reaches the maximum population of $\ket{2}$ gives us a rough estimate of the amplitude required for the $X_{12}$ pulse.
\subsection{Rough Ramsey Experiment}
The rough amplitude obtained in the rough Rabi experiment is used to perform the rough Ramsey experiment. Here, we sweep the delay between two  $\pi/2$ pulses applied at the roughly estimated $1-2 $ frequency such that the phase of the second pulse gets changed by introducing delays between the pulses to fine-tune the calibration of the qutrit $1-2$ frequency  (Fig. \ref{calibration experiments}). The frequency of fringes observed in the experiment gives us an estimate of the frequency offset from the original $1-2$ frequency.\\\\

\subsection{Fine Experiments}

Following the calibration the amplitude of the truncated Gaussian envelope and frequency $f_{12}$ of the pulse carrier that implements the $X_{12}$ and $\sqrt{X_{12}}$ gates, these are further refined by error-amplified versions of the Rabi and Ramsey experiments~\cite{Sheldon2016}. Then a derivative removal by an adiabatic gate (DRAG) correction~\cite{Motzoi2009} is applied to the quadrature component, the amplitude of which is determined by error-amplified rotary ($\sqrt{X_{12}} - \sqrt{X_{12}}^\dagger$) experiments~\cite{Sheldon2016}. Arbitrary $Z$-rotations are defined virtually by phase shifts in software~\cite{gambetta}. 

\subsection{0-1-2 discriminator error rates}
\label{discriminator}
In this subsection, we briefly explain why the limited accuracy of a custom  $0-1-2$ discriminator for the IBM backends necessitates an exponential number of circuits (in relation to the number of qutrits) that depend on the highly accurate $0-1$ discriminator for precise fidelity computation. \\

\begin{figure}
    \centering
    \includegraphics[width=0.5\linewidth]{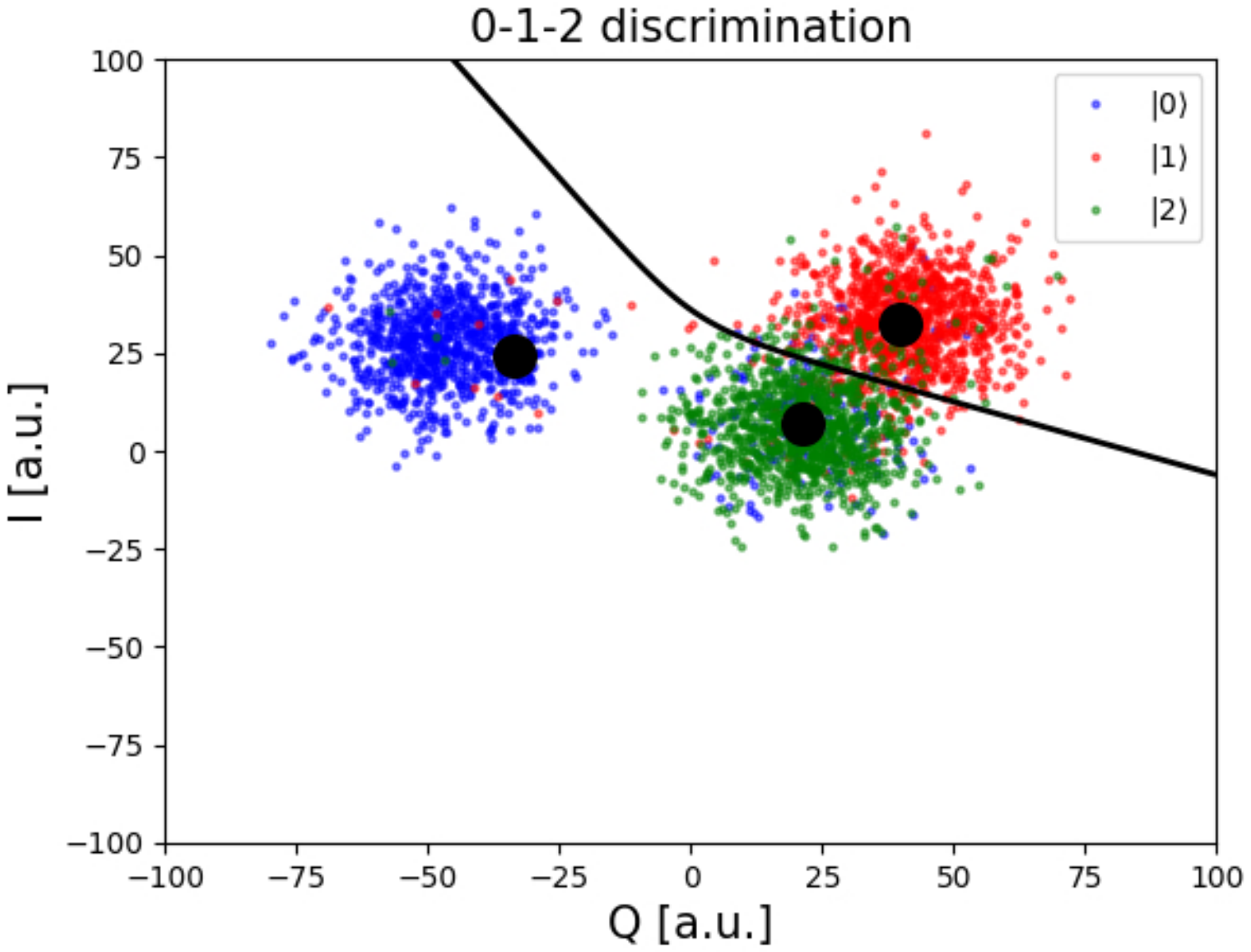}
    \caption{To discriminate between $|0\rangle$, $|1\rangle$, and $|2\rangle$ states, our model checks where the IQ point lies relative to the separatrix and classifies the point accordingly. This experiment was run on qubit 15 of \textit{ibm nazca}.}
    \label{fig:disc_nazca}
\end{figure}

We build the $0-1-2$ discriminator as follows (\cite{qiskitextbook2023,custom_disc})\\
\begin{enumerate}
\item Compute the $1\rightarrow2$ frequency through narrow-band spectroscopy.\\\\
\item Conduct the Rabi experiment to determine the $\pi$ pulse amplitude for the $1\rightarrow2$ transition. The process is as follows: First, apply a $X_{01}$ gate to transition from the $\ket{0}$ state to the $\ket{1}$ state. Next, perform a sweep of drive amplitudes at the previously obtained $1\rightarrow 2$ frequency.\\ \\
\item Construct three schedules:\
    (i) $\ket{0}$ schedule: just measure the ground state.\
    (ii) $\ket{1}$ schedule: apply a $X_{01}$ gate and measure.\
    (iii) $\ket{2}$ schedule: apply a $X_{01}$ gate, then a $X_{12}$ gate and measure their IQ (In-phase and Quadrature phase components of the measurement signal) data points.  \\\\
\item Collect data from each schedule, divide it into training and testing sets, and build a linear discriminant analysis (LDA) model for discrimination (Fig.~\ref{fig:disc_nazca}). \\
\end{enumerate}

With three centroids, the separatrix is no longer a simple line but rather a curve formed by a combination of two lines. To distinguish between the $\ket{0}$, $\ket{1}$, and $\ket{2}$ states, our model evaluates the position of the IQ point relative to the separatrix and classifies it accordingly. The LDA score (accuracy in correctly classifying the states; 1.0 is exact classification) for this discriminator is 0.873. This is very inaccurate compared to the $0-1$ discriminator, which has a readout assignment error in the range of $0.01-0.02$.

The inaccuracy described above could affect the results significantly, especially with multiple numbers of qutrits, since the error rate in correctly classifying a ternary bit string is roughly the product of error rates for individual qutrits. This motivated us to consider additional (unfortunately, exponentially scaling in system size) number of circuits where we essentially separate $\ket{1}$ from $\ket{2}$ by applying additional $X_{01}$ gates to the original circuit and using them to find the true $\ket{1}$ probabilities. It is important to emphasize that there are efforts in building an accurate $0-1-2$ discriminator (such as \cite{discrimnator}), which would rule out the requirement of additional circuits that we currently require.

\section{Obtaining ternary probability vector in the absence of perfect 0-1-2 discriminator}
\label{additional_circs}
 To benchmark the efficacy of qutrit-based encoding of the OBC AKLT states, we estimate the closeness of the probability vector of ground states from hardware and the exact probability vector by Hellinger fidelity computation (See Section \ref{definitions}). Here, the probability vector refers to the square of the absolute value of the state vector components. We restrict our analysis to Hellinger fidelity instead of the quantum fidelity (Section \ref{definitions}) because of the realness of the OBC AKLT ground states and the absence of significant phases in the state vector. This is unlike the perturbed PBC AKLT ground states, where we perform post-selection-based state tomography and phase correction for the dominant components of the state vector as they are crucial in the Berry phase estimate. \\
 
 To obtain the probability vector of the ground states from the hardware, our qutrit-based simulation requires a few additional steps. Qubit states could be measured using a highly accurate $0-1$ discriminator that classifies $\ket{0}$ state from $\ket{1}$ state with a very low readout error (See Section \ref{device_specs}). The absence of a  $0-1-2$ discriminator for the IBM Quantum backends means that both the $\ket{1}$ and $\ket{2}$ get classified as $\ket{1}$ state.  So, to compute Hellinger fidelities, one needs additional circuits to get the probability of ternary bit strings. This procedure scales exponentially in the number of qutrits, as noted in \cite{Alexey_qutrits}. In the Section \ref{discriminator}, we show typical error rates of a simple custom-built 0-1-2 discriminator \cite{custom_disc} of IBM Quantum transmons and relate it to the need for an exponential number of circuits relying only on the $0-1$ discriminator for accurate fidelity computation. However, there are other successful efforts in building  $0-1-2$ discriminator \cite{discrimnator}. Including that in our work should vastly reduce the number of circuits. The noisy simulations that we present are devoid of this discriminator-based deficiency and, hence allow us to estimate fidelities for arbitrarily large system sizes.\\

A recursive schematic of obtaining the probability vector from the hardware, which distinguishes the $\ket{1}$ states from $\ket{2}$ states by bringing them to $\ket{0}$ states, is described in Figure \ref{Fidelity_circs}. Here, the probability vector $\vec{P}_n(C_n)$ of a classical ternary $n$-bit string $((c_1,c_2,\ldots,c_n)=C_n)$ for a n-qutrit circuit is expressed in terms of the probabilities $\vec{P}_{n-1,c_n=c}(C_{n-1})$ of a classical ternary ($n-1$) bit string ($(c_1,c_2,\ldots,c_{n-1})=C_{n-1}$ and $c = 0$ or $1$) corresponding to the first ($n-1$) qutrits of the original $n$-qutrit circuit with and without $X_{01}$ gate appended on the $n$-th qutrit.  The subscript indicates that we post-select measurements where $c_n=c$ (0 or 1). \\

The base case of the recursive schematic involves computing the probabilities $\vec{P}_1$ of the $n=1$ circuit with the raw probabilities $P$  (classifying $\ket{0}$ from $\ket{1}$ and $\ket{2}$ states) straightforwardly as follows: $P_1(0) = P{(0)}$ of the single qutrit circuit; and $P_1(1) = P{(0)}$ of the same circuit with $X_{01}$ appended. Finally, with these two probabilities, one can obtain $P_1(2) = 1 - P_1(0) - P_1(1)$.\\

The recursive part of the schematic  involves the following steps:
\begin{itemize}
    \item $\vec{P}_{n}(C_{n-1}, c_n=0)= \vec{P}_{n-1,c_n=0}(C_{n-1})$ of the original circuit. 
    \item $\vec{P}_{n}(C_{n-1}, c_n=1)= \vec{P}_{n-1,c_n=0}(C_{n-1})$  of the original circuit with $X_{01}$ gate appended on the $n$-th qutrit
    \item $\vec{P}_{n}(C_{n-1},c_n=2 ) = \vec{P}_{n-1,c_n=1}(C_{n-1})$ of the original circuit $- $$\vec{ P}_{n}(C_{n-1},c_n=1)$. \\
\end{itemize}
This recursion is followed to lower and lower system sizes until our base case, which corresponds to $n=1$.

\def\myvdots{\ \vdots\ }
\begin{figure}[htbp!]
    
    \centering
     \begin{tikzpicture}
    
        \node[scale =0.8]{\begin{quantikz}[column sep =0.35 cm]
     \lstick[wires=1]{$\mathbf{(a)}$}
     &&\gate[style={fill=black!10, inner
xsep=5pt}]{Cir_1}&\meter{P_1(0) = P(0)}&
     \end{quantikz} \hspace{1.2 cm}
     \begin{quantikz}[column sep =0.35 cm]
     &\gate[style={fill=black!10, inner
xsep=5pt}]{Cir_1}&\gate[style={fill=green!40}]{X_{01}}&\meter{P_1(1) = P(0)}&
     \end{quantikz}};
    \end{tikzpicture}\\
    \begin{tikzpicture}
\node[scale=0.72] {
  \begin{quantikz}[row sep =0.4 cm, column sep =0.3 cm]
    \lstick[wires=3]{\textbf{(b)} \\\\ n qutrits}
      &  & \gate[3, nwires=2,style={fill=black!10, inner
xsep=5pt}]{Cir_{n}}&\meter{\vec{P}_{n}(C_{n-1},c_n=0) = \vec{P}_{n-1,c_n=0}(C_{n-1})}\\
      & \myvdots &\myvdots&\myvdots\\
      && &\meter{}\\
     \end{quantikz}
  \begin{quantikz}[row sep =0.4 cm, column sep =0.3 cm]
    \lstick[wires=3]{$\mathbf{(c)}$}
      &  & \gate[3, nwires=2,style={fill=black!10, inner
xsep=5pt}]{Cir_{n}}&\meter{\vec{P}_{n}(C_{n-1},c_n=1) =  \vec{P}_{n-1,c_n=0}(C_{n-1})}\\
      & \myvdots &\myvdots&\myvdots\\
      && &\gate[style={fill=green!40}]{X_{01}}&\meter{}\\
     \end{quantikz}
};
\end{tikzpicture}\\
   
    \caption{\textbf{Schematic of constructing the probability vector ($\vec{P}_n$) of qutrit circuit (b) from probabilities ($\vec{P}_{n-1,c_n=c}$ of (b) and (c)) where $c_n=\{0,1\}$:} Let $C_n$ be the sequence of classical ternary bits $c_1,c_2,\ldots,c_{n}$. Then, 
    $\vec{P}_{n}(C_{n-1},c_n=0)$ of the circuit (b) equals $ \vec{P}_{n-1,c_n=0}(C_{n-1})
    $ of the same circuit (b). $\vec{P}_{n}(C_{n-1},c_n=1)$ of  the circuit (b) equals $ \vec{P}_{n-1,c_n=0}(C_{n-1})
    $ of the circuit (c).  And 
 $\vec{P}_n(C_{n-1},c_n=2) $ of circuit (b) equals $\vec{P}_{n-1,c_n=1}(C_{n-1})$
of circuit (b) $-  $$\vec{P}_{n}(C_{n-1},c_n=1)$. This recursion on the number of qutrits is followed until the circuit has just one qutrit. At the level of one qutrit, in terms of raw probabilities $P$, $P_1(0) = P(0)$ of the first circuit in (a) and $P(1) = P(0)$ of the second circuit in (a) and $P_1(2) = 1- P_1(0) - P_1(1)$.}
 \label{Fidelity_circs}
\end{figure}
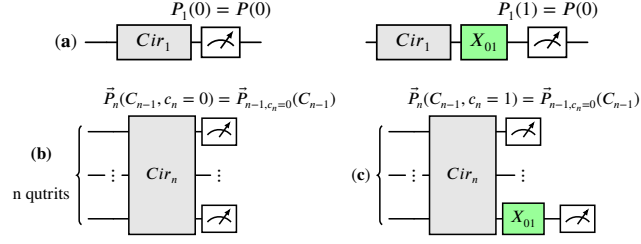

\section{Pulse parameters and gate fidelities}
Table \ref{tab:gate_fids} contains the pulse parameters obtained post-calibration and their associated Error Per Clifford (EPC).
\begin{table}[h!]
\caption{Qutrit gate fidelities  (in Error Per Clifford (EPC)) and calibrated parameters of the pulse gates.}
    \begin{center}
    \begin{tabular}{|c|c|c|c|l|} \hline 
         Qutrit&  $X_{12}$ amplitude&  Anharmonicity& Beta ($\beta$) &$X_{12}$ gate error (EPC/2)\\ \hline 
         1&  1.486e-01&  -3.4199e+08&  2.6827e-01&0.00038 $\pm$ 0.00022\\ \hline 
         2&  1.6537e-01&  -3.3948e+08&  -1.3157&0.00035 $\pm$ 0.00005\\ \hline
 3& 2.0118e-01& -3.4274e+08& -3.0499e-01&0.00027 $\pm$ 0.00005\\\hline
    \end{tabular}
    \end{center}
    \caption{For $n=2$ (both OBC and PBC  (Fig.~\ref{fig:obc_hardware} A and Fig.~\ref{PBC_hardware} C in the main text), qutrits 1 and 2 were used. For $n=3$ (Fig.~\ref{fig:obc_hardware} B in the main text), qutrits 1, 2, and 3 were used. For $n=4$ (Fig.~\ref{fig:obc_hardware} C in the main text), qutrits 1, 2, 3, and qubit 4 were used (one does not need to calibrate qutrit 4 for the state preparations of interest). Qubits 3, 2, 1, and 4 exist in a single linear chain in that order.}
    \label{tab:gate_fids}
\end{table}
\newpage
\section{Parameters for the circuits (OBC and PBC) run on hardware }
\label{github}
Circuit diagrams with pulse gates (and $0-1$ manifold gates) along with the states that they produce, are shown below. We refer the reader to the GitHub files (\href{https://github.com/keerthikumara/Spin-1}{GitHub repository}) for precise parameter values.\\
\begin{figure}[!h]
        \begin{center}
                \includegraphics[height=0.2\textheight]{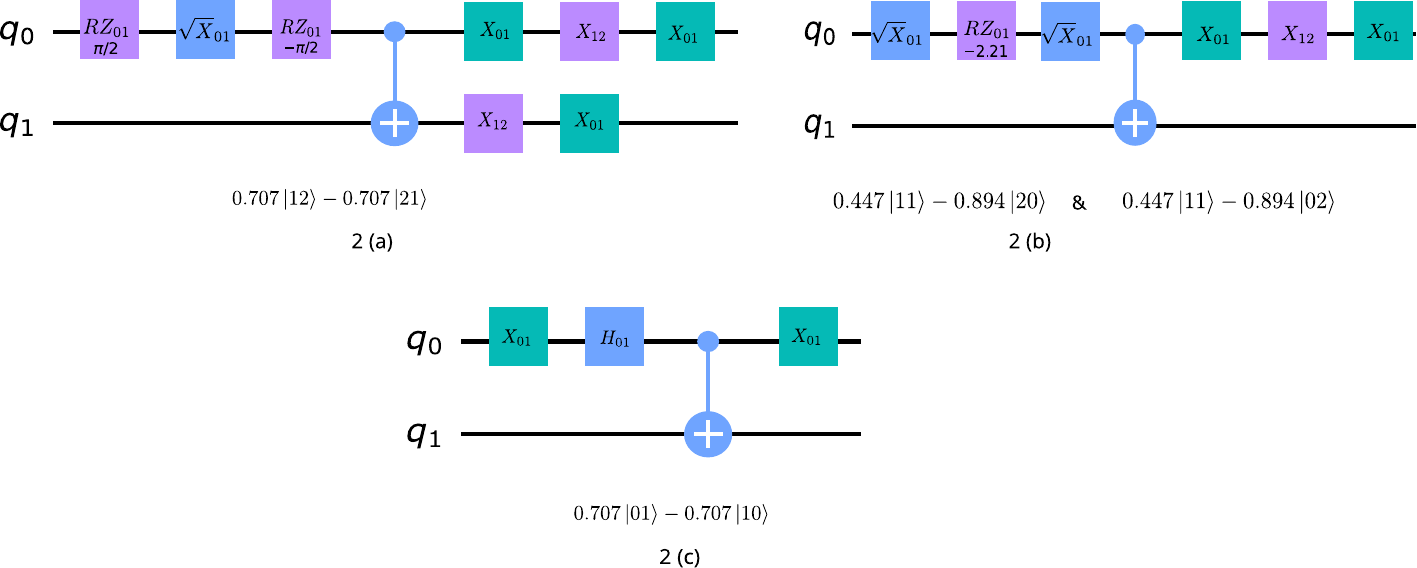}
        \end{center}
            
        \caption{OBC AKLT ground-state circuits for $n=2$, 2 a, 2 b in Fig.  \ref{fig:obc_hardware} D in the main text. 2(a) captures one of the ground states. 2(b) captures two out of four ground states up to an inversion in the qutrit connectivity list. 2(c) captures the fourth ground state, but it can be prepared in $0-1$ manifold without any qutrit gates and hence can be obtained with $99-100 \%$ Hellinger fidelity. }
        \label{gs_2}
    \end{figure}
The reasoning behind the exceptionally high-fidelity state (2 c) is that it lies entirely within the default 0-1 manifold, where these transmons typically function. 
It is important to note that, given these high probabilities, they may occasionally exceed 100\% (e.g., 100.1\%). This occurs because error mitigation techniques can produce corrected values that are not strictly bounded between 0 and 1, unlike raw probabilities. 
    \begin{figure}[!h]
        \begin{center}
            \includegraphics[height=0.3\textheight]{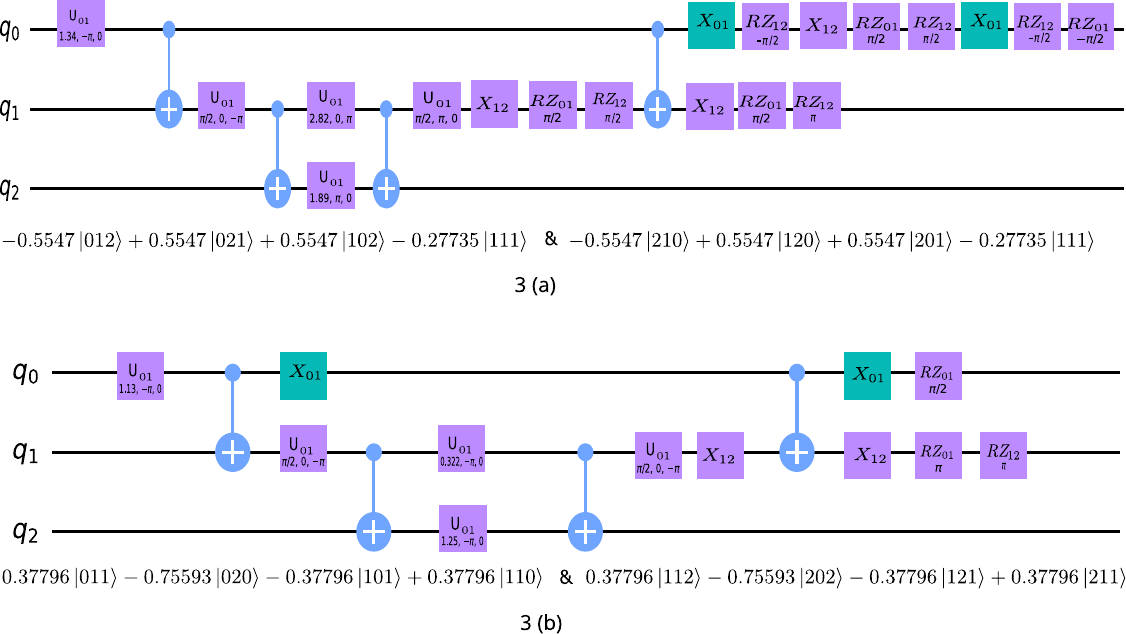}
        \end{center}
        
        \caption{OBC AKLT ground-state circuits for $n=3$, 3 a, 3 b in Fig. \ref{fig:obc_hardware} D in the main text.  3(a) captures two out of four ground states up to an inversion in the qutrit connectivity list.  3(b) captures two out of four ground states up to a trivial layer of single qutrit gates taking $\ket{0} \leftrightarrow \ket{2}$. For Hellinger fidelity computation, the same set of circuits used for computing the full probability vector of 3(b) suffices to compute the full probability vector of the 3(b) with $\ket{0} \leftrightarrow \ket{2}$ transformation in the end. }
        \label{gs_3}
    \end{figure}
\begin{figure}[!h]
        \begin{center}
            \includegraphics[height=0.3\textheight]{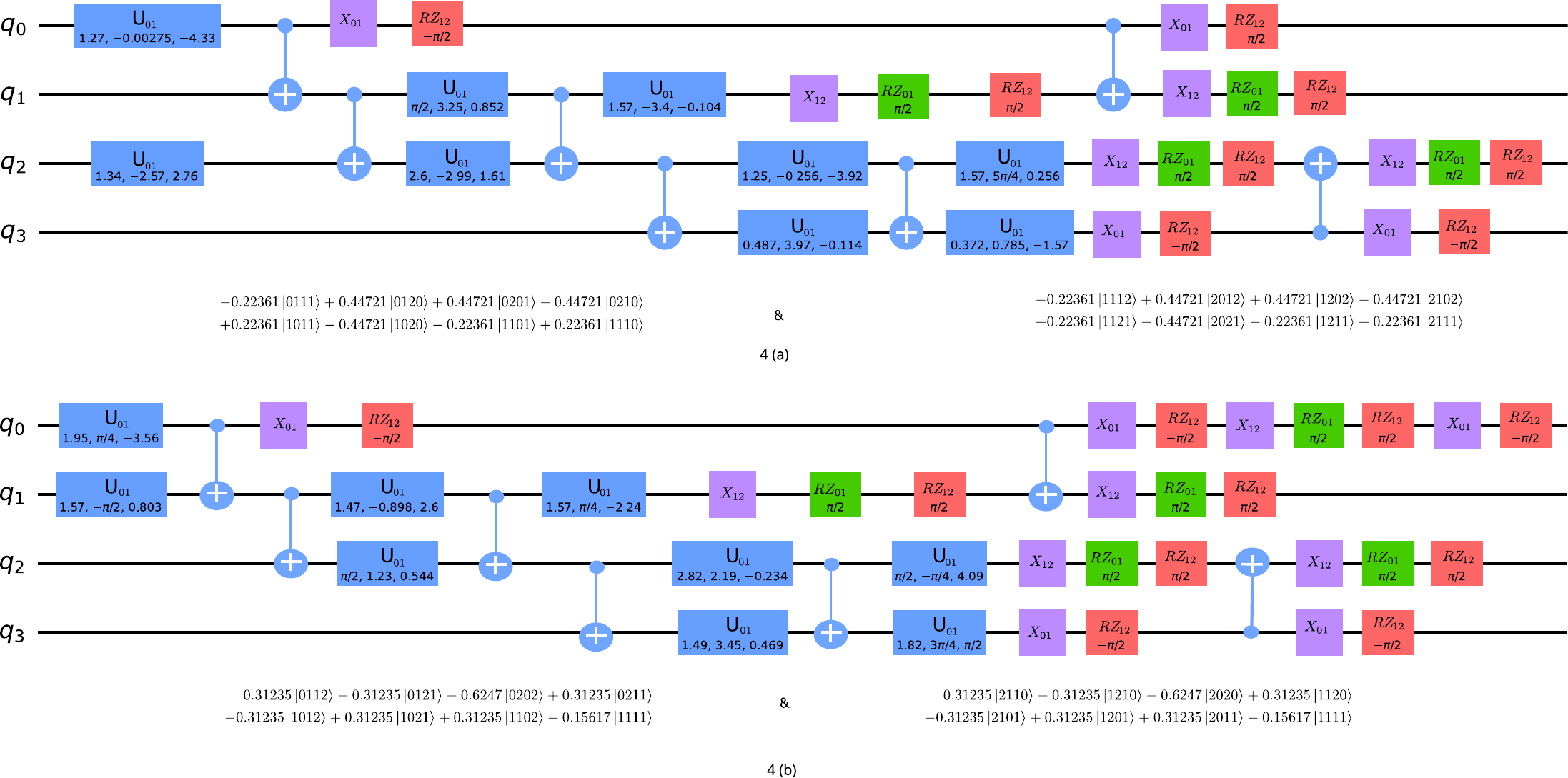} 
        \end{center}
       
        \caption{OBC AKLT ground-state circuits for $n=4$, 4 a, 4 b in Fig. \ref{fig:obc_hardware} D in the main text. 4(a) captures two out of four ground states up to a trivial layer of single qutrit gates taking $\ket{0} \leftrightarrow \ket{2}$. For Hellinger fidelity computation, the same set of circuits used for computing the full probability vector of 4(a) suffices to compute the full probability vector of the 4(a) with $\ket{0} \leftrightarrow \ket{2}$ transformation in the end.  4(b) captures the other two ground states up to an inversion in the qutrit connectivity list. }
        \label{gs_4}
    \end{figure}
    \begin{figure}[!h]
        \begin{center}
           \includegraphics[height=0.1\textwidth]{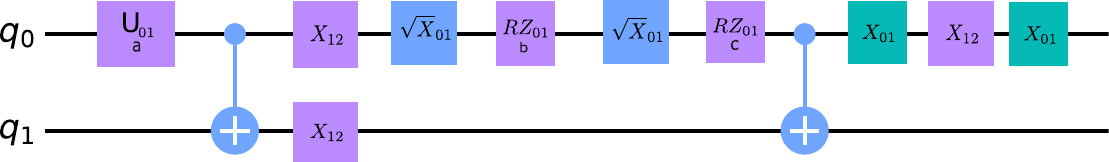} 
        \end{center}
        
        \caption{PBC perturbed AKLT ground-state circuits. The $U_{01}$ gate is the standard $U$ gate used in qubits-based circuits \ref{gates}. The parameters $(a,b,c)$ depend on the value of perturbation and can be found in the files in the Github repository attached.}
        \label{pbc_circ}
    \end{figure}

\section{Phase differences between dominant vector components \label{phase_experiments}}
\subsubsection{Phase differences between the ground state vector components}
\label{ac_stark_shifts}
Single qutrit gates produce ac-Stark shifts themselves, in addition to the non-negligible Stark phase shift introduced by the CNOT gate when the control qutrit is in $\ket{2}$ state. This has been noted in \cite{galda2021implementing, Alexey_qutrits}. Empirically, the phase difference between the $\ket{02}$ and $\ket{20}$ components of the state produced by the real hardware circuit is $\;\text{Arg}[\alpha / (\beta e^{i \phi_s})]$ where $\phi_s$ is the contribution due to the Stark shift. We need to correct these additional $\phi_s$ by performing a phase shift experiment by adding controllable phase differences between the state vector components. This experiment will be used to correct the unwanted phase accumulation from the Stark shift. This method was previously implemented in \cite{galda2021implementing} and works best for smaller systems like the one we will discuss. For general spin-1 systems, it is essential to construct a high-fidelity entangling gate like the one being studied in \cite{entangling_qutrits}. In the following section, we present simplified noisy simulations of fidelities from scalable ansatz of the ground states, which would be more suitable when the effect of noise sources such as charge dispersion and ac-Stark shifts due to higher levels can be minimized in future devices.

\subsubsection{Phase shift experiments}
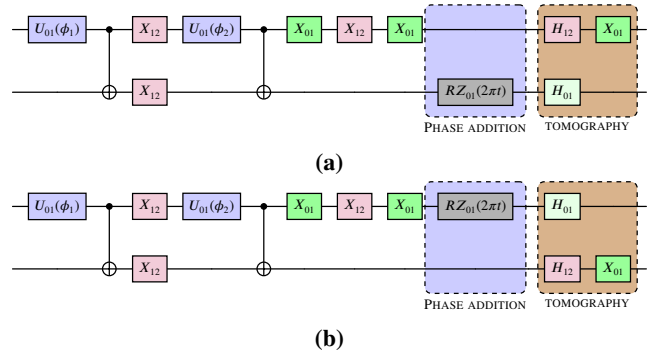
\begin{figure}[htbp!]
\begin{center}

    \begin{tikzpicture}
        \node[scale =0.6]{\begin{quantikz}[row sep = 0.8 cm,column sep =0.35 cm]
           & \gate[style={fill=blue!20}]{U_{01}(\phi_1)} &
\ctrl{1}&\gate[style={fill=purple!20}]{X_{12}}&\gate[style={fill=blue!20}]{U_{01}(\phi_2)}&\ctrl{1}&\gate[style={fill=green!40}]{X_{01}}&\gate[style={fill=purple!20}]{X_{12}}&\gate[style={fill=green!40}]{X_{01}}&\gategroup[2,steps=1,style={dashed,rounded
corners,fill=blue!20, inner
xsep=5pt},background,label style={label
position=below,anchor=north,yshift=-0.2cm}]{{\sc
Phase addition}} & &\gate[style={fill=purple!20}]{H_{12}}\gategroup[2,steps=2,style={dashed,rounded
corners,fill=brown!60, inner
xsep=1pt},background,label style={label
position=below,anchor=north,yshift=-0.2cm}]{{\sc tomography}}&\gate[style={fill=green!40}]{X_{01}}&  \\
& &\targ{} &\gate[style={fill=purple!20}]{X_{12}}&& \targ{} &&&&\gate[style={fill=black!30}]{RZ_{01}(2\pi t)}&& \gate[style={fill=green!10}]{H_{01}} &&
\end{quantikz}
};
    \end{tikzpicture}\\
    \textbf{(a)}\\
    \begin{tikzpicture}
        \node[scale =0.6]{\begin{quantikz}[row sep = 0.8 cm,column sep =0.35 cm]
           & \gate[style={fill=blue!20}]{U_{01}(\phi_1)} &
\ctrl{1}&\gate[style={fill=purple!20}]{X_{12}}&\gate[style={fill=blue!20}]{U_{01}(\phi_2)}&\ctrl{1}&\gate[style={fill=green!40}]{X_{01}}&\gate[style={fill=purple!20}]{X_{12}}&\gate[style={fill=green!40}]{X_{01}}&\gate[style={fill=black!30}]{RZ_{01}(2\pi t)}\gategroup[2,steps=1,style={dashed,rounded
corners,fill=blue!20, inner
xsep=5pt},background,label style={label
position=below,anchor=north,yshift=-0.2cm}]{{\sc
Phase addition}} & &\gate[style={fill=green!10}]{H_{01}}\gategroup[2,steps=2,style={dashed,rounded
corners,fill=brown!60, inner
xsep=1pt},background,label style={label
position=below,anchor=north,yshift=-0.2cm}]{{\sc tomography}}&&  \\
& &\targ{} &\gate[style={fill=purple!20}]{X_{12}}&& \targ{} &&&&&&\gate[style={fill=purple!20}]{H_{12}}& \gate[style={fill=green!40}]{X_{01}} &
\end{quantikz}
};
    \end{tikzpicture}\\
    \textbf{(b)}
     \end{center}
     \caption{\textbf{Phase shift experiments:} We perform phase shift experiment by introducing a controllable phase $2 \pi t$ ($t$ $\in $ [0,1]) between the $a_{02}$ and $a_{11}$ components in the circuit (a) and the $a_{20}$ and $a_{11}$ components in the circuit (b). Then state tomography analog of qutrits is performed, resulting in the output $P(00) = \frac{|a_{02} - a_{11} e^{(i \phi + i \phi_s)}|}{4}$ for the circuit (a) and  $P(00) = \frac{|a_{20}- a_{11} e^{(i \phi + i \phi_s)}|}{4}$ for the circuit (b). In the figure, $\phi$ is the controllable phase and $\phi_s$ is the contribution from the Stark shift, which we wish to correct.}  \label{phase_rot}
    
\end{figure}

There are studies regarding the general state tomography approach for qutrit circuits \cite{qutrittomo1,qutrittomo2}. These approaches must differ significantly from qubit state tomography because the number of generators of the SU(3) group (acting on a qutrit) is 8 and would require around $8^N$ circuits to fully perform state tomography on an $N$ qutrit circuit (compare this with $3^N$ circuits required for an $N$ qubit state tomography). However, since we know the form of the states Eq.~\ref{PBC_state}, we can focus on just two sets of tomography circuits Fig. \ref{phase_rot}. These two circuits can be used to compute the (real) phase difference between the ($\ket{02}$ and $\ket{11}$) components and ($\ket{20}$ and $\ket{11}$). Let the coefficient of the $\ket{ij}$ component be  $a_{ij}$. Then, the output probability of $\ket{00}$ state of the circuit (a) in Fig.~\ref{phase_rot} is
\begin{equation}
    P(\ket{00}) = \frac{|(a_{02} - a_{11} e^{i \phi_s})|^2}{4} \ ,
\end{equation}
where $\phi_s$ is the contribution due to the Stark shifts described earlier.\\

For circuit (b) in Fig. \ref{phase_rot}
\begin{equation}
    P(\ket{00}) = \frac{|(a_{20} - a_{11} e^{i \phi_s})|^2}{4} 
\end{equation}
There is no known straightforward relationship between the phase additions in both of these circuits, we correct them separately.\\

\begin{center}
        \begin{figure}[!h]
        \begin{center}

        \includegraphics[height=0.7\textheight]{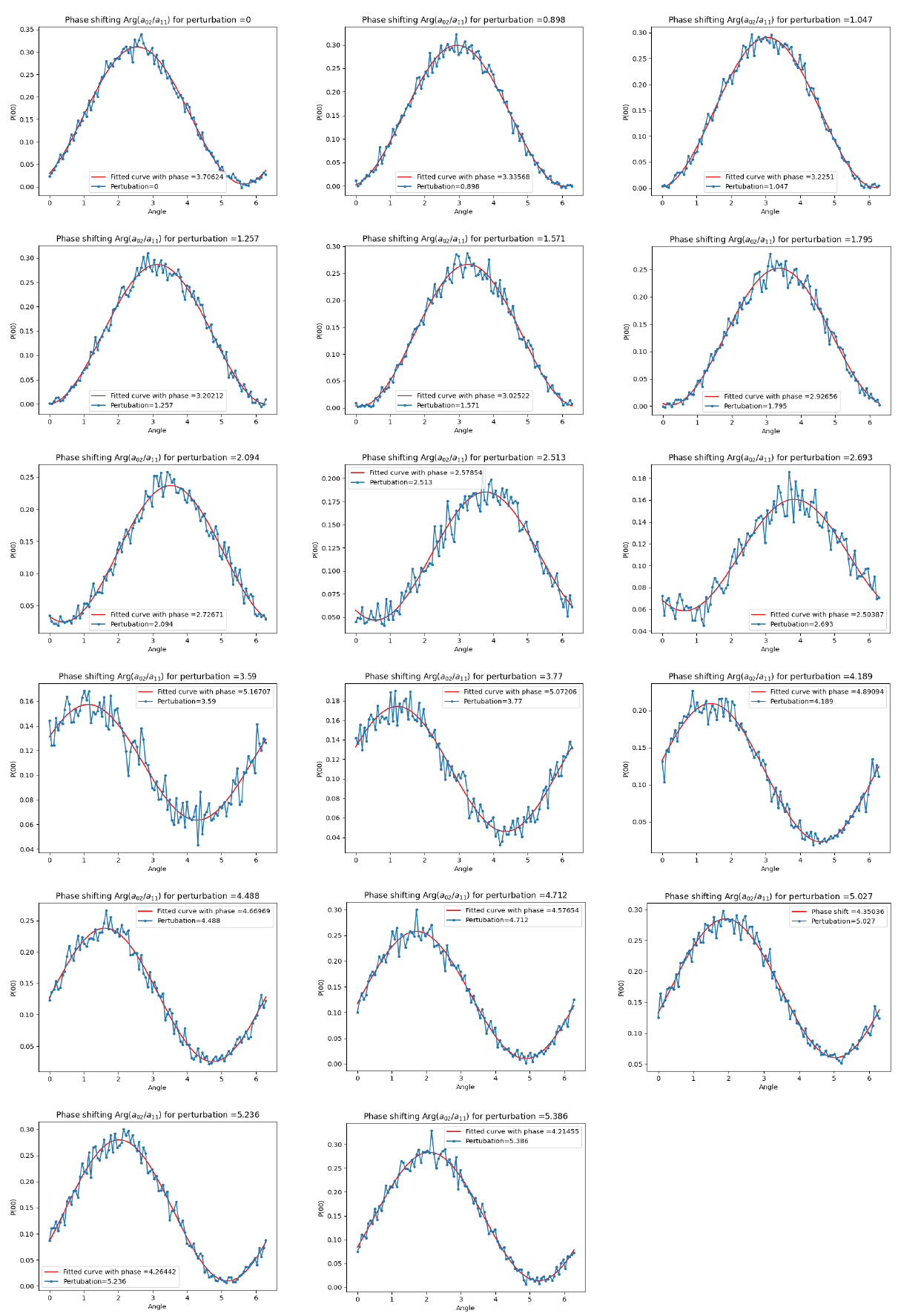}

        \end{center}
        \caption{\textbf{Phase shift experiments 1:} Phase shift plots corresponding to ($\ket{02},\ket{11}$) states for all the perturbed ground states (perturbation angle shown in the subplots) are considered in this work. }
        \label{phase_shift_experiments1}
    \end{figure}
    \end{center}

\begin{center}
        \begin{figure}[!h]
        \begin{center}
            \includegraphics[height=0.8\textheight]{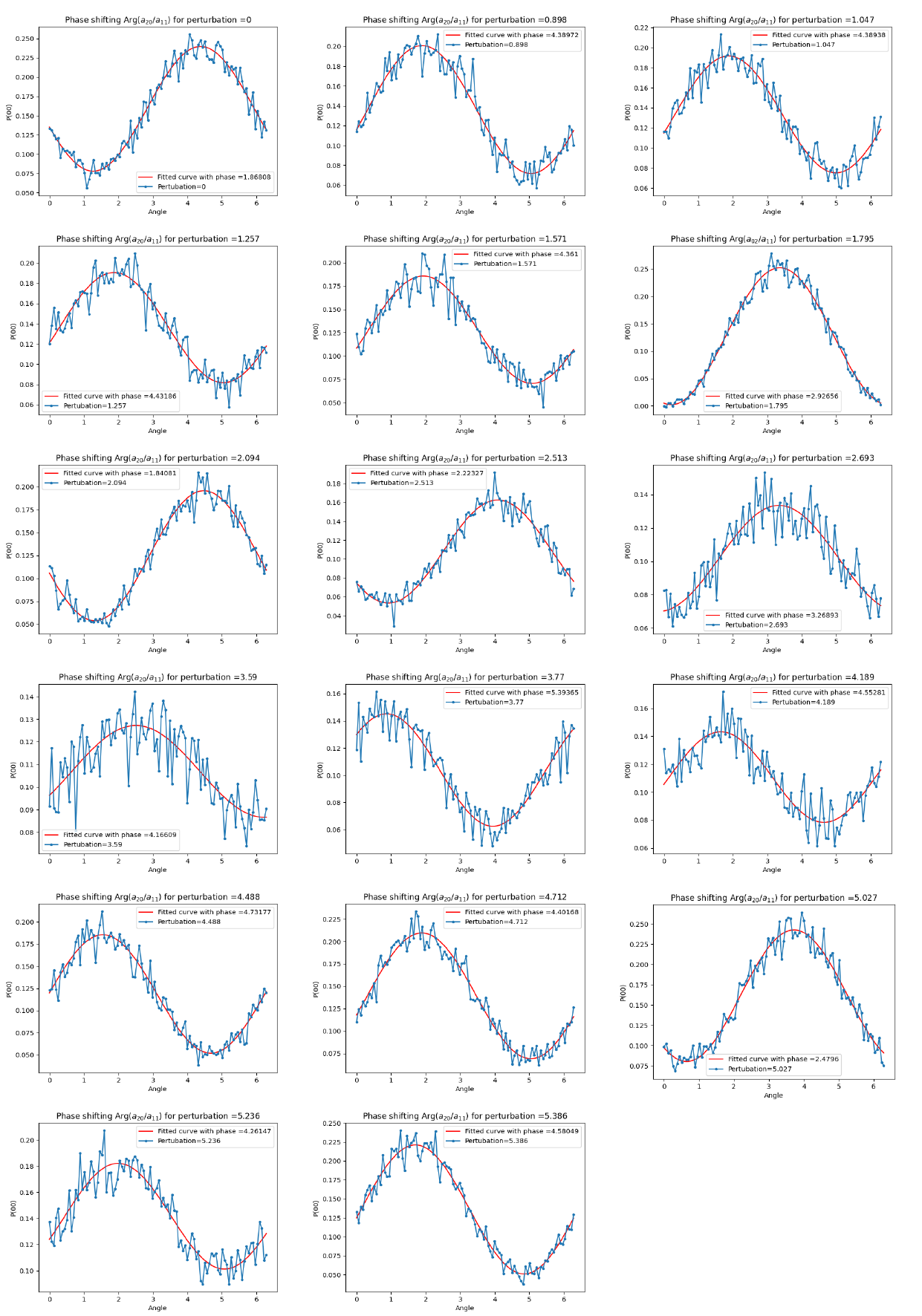}
        \end{center}
        
        \caption{\textbf{Phase shift experiments 2:} Phase shift plots corresponding to ($\ket{20},\ket{11}$) states for all the perturbed ground states (perturbation angle shown in the subplots) are considered in this work.}
        \label{phase_shift_experiments2}
    \end{figure}
    \end{center}

The first set of phase shift experiments was performed by introducing a controllable phase $2 \pi t$ ($t$ $\in $ [0,1]) between the $\ket{02}$  and $\ket{11}$ components in the circuit (a) of Fig.~\ref{phase_rot} resulting in plots shown in Fig.\ref{phase_shift_experiments1}. Then state tomography analog of qutrits is performed, resulting in the output $P(00) = \frac{|a_{02} - a_{11} e^{(i \phi + i \phi_s)}|}{4}$ for the circuit (a). In Fig.~\ref{phase_shift_experiments1}, we see that the peak occurs at the point where the phase difference between the  $\ket{02}$, $\ket{11}$ components gets canceled by the controlled rotation $\phi$ and informs us the required phase correction (Fig.~\ref{phase_corr})  for our state. 
\newpage
 The second set of phase shift experiments was performed by introducing a controllable phase $2 \pi t$ ($t$ $\in $ [0,1]) between the $\ket{20}$ and $\ket{11}$ components in the circuit (b) of Fig.~\ref{phase_rot}. Then state tomography is performed, resulting in the output $P(00) = \frac{|a_{20} - a_{11} e^{(i \phi + i \phi_s)}|}{4}$ for the circuit (b). In Fig.~\ref{phase_shift_experiments2}, we see that the peak occurs at the point where the phase difference between the  $\ket{20}$, $\ket{11}$ components gets canceled by the controlled rotation $\phi$ and informs us about the exact phase correction (Fig.~\ref{phase_corr}) required for our state.
    
It is important to note that the phase shift plots need not necessarily reach their theoretical maximum and minimum owing to  imperfect final states (Fidelity < 100 $\%$) prepared in the hardware. The hardware data corresponding to these experiments can be found in the \href{https://github.com/keerthikumara/Spin-1}{GitHub repository} attached.
 
    \begin{figure}
        \begin{center}
             \includegraphics[width=1\linewidth]{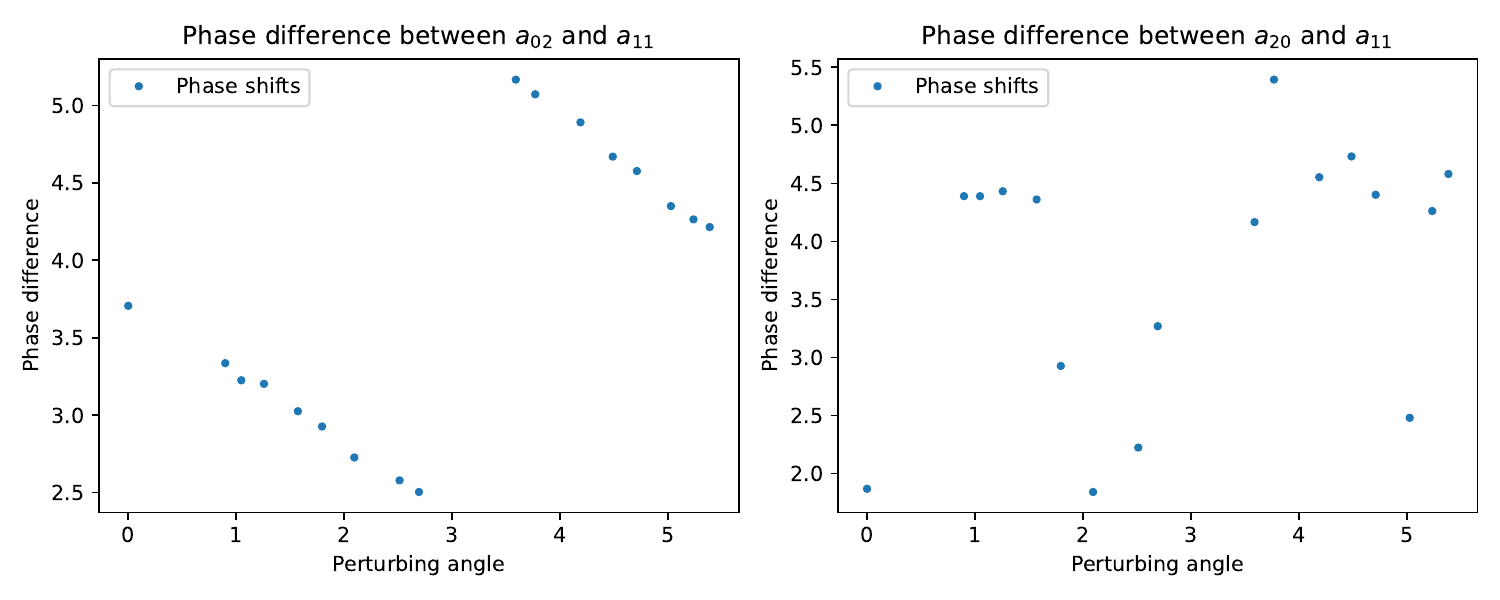}
        \end{center}
       
        \caption{The phase difference between $\ket{20}$ or $\ket{02}$ and $\ket{11}$ components. (a) The phase difference between $\ket{02}$ and $\ket{11}$ components obey a very regular trend for various controlled angle rotations. (b) The phase difference between $\ket{20}$ and $\ket{11}$ components depends on the device conditions and fluctuates significantly. Hence, care has to be taken to run these circuits within a single session to get the phase corrections.}
        \label{Combined_phase_shifts}
    \end{figure}

\subsubsection{Phase correction}

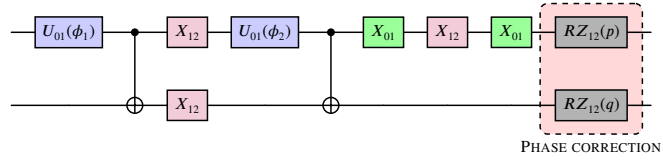
\begin{figure}[htbp!]
\begin{center}
    \begin{tikzpicture}
        \node[scale =0.7]{\begin{quantikz}[row sep = 0.8 cm,column sep =0.45 cm]
           & \gate[style={fill=blue!20}]{U_{01}(\phi_1)} &
\ctrl{1}&\gate[style={fill=purple!20}]{X_{12}}&\gate[style={fill=blue!20}]{U_{01}(\phi_2)}&\ctrl{1}&\gate[style={fill=green!40}]{X_{01}}&\gate[style={fill=purple!20}]{X_{12}}&\gate[style={fill=green!40}]{X_{01}}&\gate[style={fill=black!30}]{RZ_{12}(p)}\gategroup[2,steps=1,style={dashed,rounded
corners,fill=pink!60, inner
xsep=5pt},background,label style={label
position=below,anchor=north,yshift=-0.2cm}]{{\sc
Phase correction}} & \\
& &\targ{} &\gate[style={fill=purple!20}]{X_{12}}&& \targ{} &&&&\gate[style={fill=black!30}]{RZ_{12}(q)}&
\end{quantikz}
};
    \end{tikzpicture}\\  
    \end{center}
     \caption{\textbf{Phase correction:} The Fig. \ref{phase_rot} provides the phase difference between $\ket{02}$ and $\ket{11}$ components (circuit(a)) and $\ket{20}$ and $\ket{11}$ (circuit (b)). Using a layer of $RZ_{12}$ gates, one can introduce the expected phase difference between those components. This correction procedure is noiseless as the added gates in this procedure are all virtual noiseless $RZ$ rotations.}  \label{phase_corr}
    
\end{figure}

For the berry phase computation, the phase shift experiments need to be done for all the perturbing angles corresponding to the partitions we use (See Fig. \ref{phase_shift_experiments1}, \ref{phase_shift_experiments2}). The angle at which the $P(\ket{00})$ of the circuit (a or b) is maximum corresponds to the point where $\ket{02}$ or $\ket{20}$ and $\ket{11}$ are in phase. These experiments give us the phase differences that already exist for the state in the absence of the controllable phase addition. Based on the obtained phase differences, one can add a layer of $RZ_{12}$ gates to add the expected phase to our state Fig.~\ref{phase_corr}. This procedure does not affect the fidelities of our ansatz circuit because the additional $RZ$ gates are all zero duration virtual gates \cite{gambetta} and hence noiseless.

\section{Compilation of matrix product states}
\label{circuit_compilation}

\def\myvdots{\ \vdots\ }
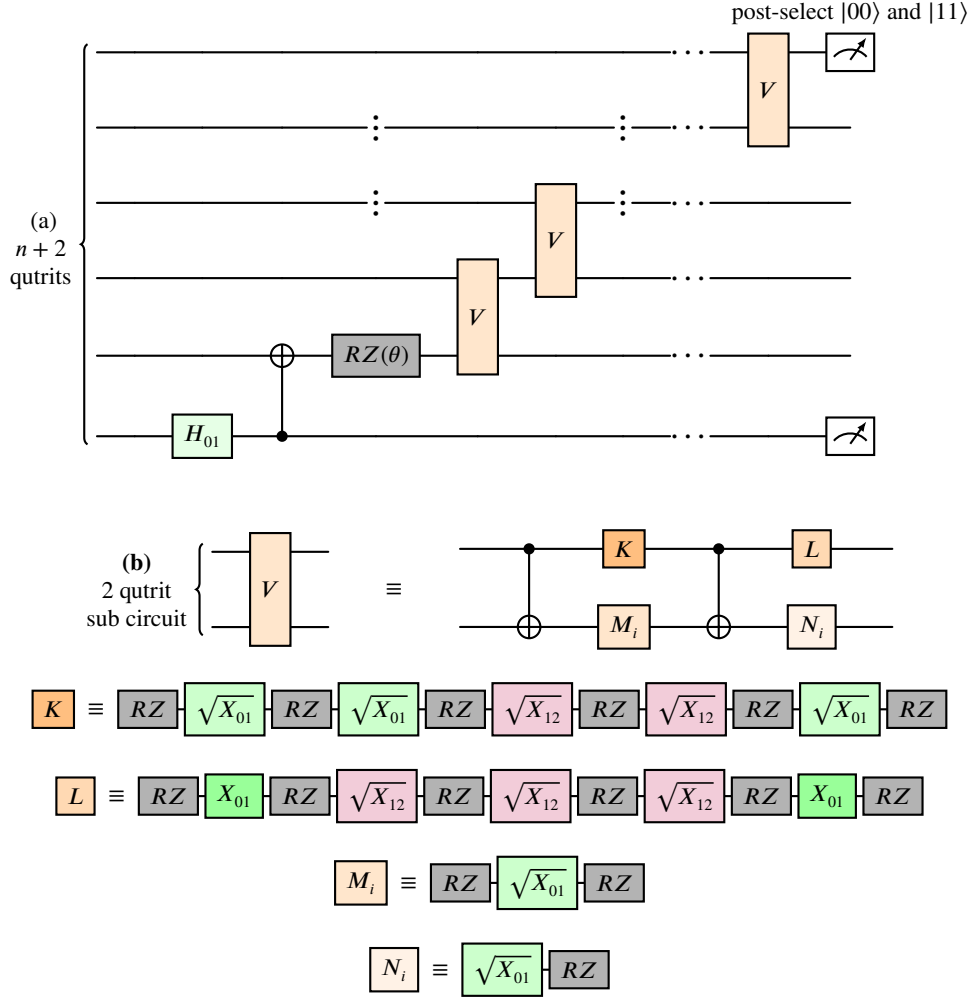
\begin{figure}[htbp!]
    \centering
    \begin{tikzpicture}
\node[scale=1.0] {\begin{quantikz}
  \lstick[wires=6]{\\\ (a) \\ $n +2$ \\ qutrits}
&&& &&&&&\scalebox{1.5}{\ldots}&\gate[2, style={fill=orange!20}]{V}&\meter{\text{post-select $\ket{00}$ and $\ket{11}$}}\\
&&&&\scalebox{1.5}{\vdots} &&&\scalebox{1.5}{\vdots}&\scalebox{1.5}{\ldots}&&\\
&&& &\scalebox{1.5}{\vdots}&&\gate[2, style={fill=orange!20}]{V}&\scalebox{1.5}{\vdots}&\scalebox{1.5}{\ldots}&&\\
 && &&&\gate[2, style={fill=orange!20}]{V}&&&\scalebox{1.5}{\ldots}&&\\
&& &\targ{}& \gate[style={fill=black!30}]{RZ{(\theta)}}&&&&\scalebox{1.5}{\ldots}&&\\
& &\gate[style={fill=green!10}]{H_{01}}& \ctrl{-1}&&&&&\scalebox{1.5}{\ldots}&&\meter{}\\
\end{quantikz}
};
\end{tikzpicture}\\
    \begin{tikzpicture}
     \node[scale=1.0]{
        \begin{quantikz}
     \lstick[wires=2]{\textbf{(b) }\\ 2 qutrit \\ sub circuit}
     &\gate[2,style={fill=orange!20}]{V}&\\
     &&
     \end{quantikz}
      \hspace{0.5 cm}
      $\equiv$
      \hspace{0.5 cm}
    \begin{quantikz}[column sep =0.75 cm]
     &\ctrl{1}&\gate[style={fill=orange!50}]{K}&\ctrl{1}&\gate[style={fill=orange!30}]{L}&\\
     &\targ{}&\gate[style={fill=orange!20}]{M_i}&\targ{}&\gate[style={fill=orange!10}]{N_i}&
     \end{quantikz}
      };
      \end{tikzpicture}\\
        \begin{tikzpicture}
     \node[scale=1.0]{
        \begin{quantikz}
     \gate[style={fill=orange!50}]{K}
     \end{quantikz}
      $\equiv$
    \begin{quantikz}[column sep =0.1 cm]
      \gate[style={fill=black!30}]{RZ}&\gate[style={fill=green!20}]{\sqrt{X_{01}}}&\gate[style={fill=black!30}]{RZ}&\gate[style={fill=green!20}]{\sqrt{X_{01}}}&\gate[style={fill=black!30}]{RZ}&\gate[style={fill=purple!20}]{\sqrt{X_{12}}}&\gate[style={fill=black!30}]{RZ}&\gate[style={fill=purple!20}]{\sqrt{X_{12}}}&\gate[style={fill=black!30}]{RZ}&\gate[style={fill=green!20}]{\sqrt{X_{01}}}&\gate[style={fill=black!30}]{RZ}
     \end{quantikz}
      };
    \end{tikzpicture}\\
     \begin{tikzpicture}
     \node[scale=1.0]{
        \begin{quantikz}
     \gate[style={fill=orange!30}]{L}
     \end{quantikz}
      $\equiv$
    \begin{quantikz}[column sep =0.1 cm]
      \gate[style={fill=black!30}]{RZ}&\gate[style={fill=green!40}]{{X_{01}}}&\gate[style={fill=black!30}]{RZ}&\gate[style={fill=purple!20}]{\sqrt{X_{12}}}&\gate[style={fill=black!30}]{RZ}&\gate[style={fill=purple!20}]{\sqrt{X_{12}}}&\gate[style={fill=black!30}]{RZ}&\gate[style={fill=purple!20}]{\sqrt{X_{12}}}&\gate[style={fill=black!30}]{RZ}&\gate[style={fill=green!40}]{{X_{01}}}&\gate[style={fill=black!30}]{RZ}
     \end{quantikz}
      };
    \end{tikzpicture}\\
     \begin{tikzpicture}
     \node[scale=1.0]{
        \begin{quantikz}
     \gate[style={fill=orange!20}]{M_i}
     \end{quantikz}
      $\equiv$
    \begin{quantikz}[column sep =0.1 cm]
      \gate[style={fill=black!30}]{RZ}&\gate[style={fill=green!20}]{\sqrt{X_{01}}}&\gate[style={fill=black!30}]{RZ}
     \end{quantikz}
      };
    \end{tikzpicture}\\
     \begin{tikzpicture}
     \node[scale=1.0]{
        \begin{quantikz}
     \gate[style={fill=orange!10}]{N_i}
     \end{quantikz}
      $\equiv$
    \begin{quantikz}[column sep =0.1 cm]
      \gate[style={fill=green!20}]{\sqrt{X_{01}}}&\gate[style={fill=black!30}]{RZ}
     \end{quantikz}
      };
    \end{tikzpicture}
    \caption{\textbf{Compilation of PBC AKLT ground state}. The compilation shown is in terms of native qutrit gates. An equivalent compilation exists in terms of qubit gates.}
    \label{OBC_PBC_Ansatz}
\end{figure}

To derive scalable circuits to prepare AKLT ground states, we use the encoding of MPS into circuits described in \cite{mps_encoding}. For the PBC AKLT model with perturbation angle, $\theta$, the circuit is shown in Fig. \ref{OBC_PBC_Ansatz}. The two ancillary qutrits at the edges are measured, and their outcome is post-selected to generate the correct classical correlations coming from the periodic boundary. \\

Note that this method generates a circuit of entangling isometries, denoted by $V$ in Fig. \ref{OBC_PBC_Ansatz}. Each isometry must then be compiled to the native gate set on a device. One way to do this is by embedding the isometry into a unitary and then performing full unitary compilation. However, for a given isometry, there is no canonical choice of embedding, and different embeddings can lead to circuits with vastly different gate counts and, therefore, amounts of noise. \\

Instead, we find noise-efficient compilations by optimizing an ansatz such that it matches the action of the isometry on the relevant subspace. This is akin to solving a multi-state preparation problem and avoids full unitary compilation. The cost function we minimize is as follows: 
\begin{align}
    C(U) &= \sum_j \left\lvert\lvert U\ket{j} - V\ket{j}\rvert\right\rvert, \label{cost_f}
\end{align}
where $U$ represents an ansatz circuit, $V$ the target isometry, and $\{\ket{j}\}$ can be generated by identifying the possible states of the input wires of $V$ in the circuit. In Fig. \ref{OBC_PBC_Ansatz}, the relevant actions of $V$ that we optimize for are: 
\begin{align}
    V\ket{00} &= -\sqrt{\frac{1}{3}} \ket{01} - \sqrt{\frac{2}{3}} \ket{12}, \\ 
    V\ket{01} &= \sqrt{\frac{2}{3}} \ket{00} + \sqrt{\frac{1}{3}} \ket{11}
\end{align}

For qutrits, the native gate set is described in Section \ref{gates}, while for qubits we use the standard CNOT + ZSX gate set available on IBM devices. In both cases, our ansatz circuit is built from CNOT gates, with each CNOT dressed with parametrized single site gates on its four wires. We optimize over all possible structures of CNOT gates, starting with structures with the fewest CNOTs. For a given structure, we optimize the single site gates to get the lowest possible value of the cost function. This continues until a structure is found for which the cost function converges below a chosen tolerance.   \\

The optimization of the single site gates can be performed by explicitly parametrizing them and using algorithms like gradient descent or BFGS. In this work, we use the environment tensor method \cite{evenbly_vidal} to optimize the gates. Once a solution has been found, the generic single site gates are decomposed into pulses available on the device.  \\ 

This method is also used to compile the gates necessary for Berry phase computation (See Section \ref{gen_hadamard_test}). We identify the relevant input states based on the measurement probabilities required for the Hadamard test and ensure that our compilation preserves these probabilities. Since the cost function in Eq. \ref{cost_f} is sensitive to phases, it is suitable for compiling the Hadamard test circuit. \\

\section{Modeling and simulating noise for transmon qutrits}
\label{noise_model}

A superconducting qutrit interacting with its environment will thermalize. This thermalization can take place either with or without a loss of energy. In the first case, the qutrit will spontaneously decay from a higher energy state to a lower one, giving off a quantum of energy to the environment. Since the environment has, by definition, many more degrees of freedom than the qutrit, this corresponds to a macrostate with a much larger set of microstates. The resulting increase in entropy can be interpreted as thermalization. We model this behavior with the amplitude damping channel (\ref{amp_damp}). In the second case, the off-diagonal elements of the density matrix will decay while the diagonal elements stay preserved. Therefore, the energy of the qutrit will remain unchanged, but its von Neumann entropy will increase, once again allowing it to thermalize. The dephasing channel (\ref{dephasing}) is used to model this phenomenon.\\\\

We will now heuristically derive the Kraus operators of each channel and then relate their free parameters to experimentally observable  $T_1$ and $T_2$ times. 

\subsection{Amplitude damping channel}
\label{amp_damp}
Amplitude damping of qutrits corresponds to a unitary on the qutrit + environment system with the following action: \\

\begin{eqnarray}
      U\ket{0_q}\lvert 0_E\rangle = \ket{0_q}\ket{0_E}\\
    U\ket{1_q}\ket{0_E} = \sqrt{p_{\Delta_{0-1}}}\ket{0_q}\ket{\Delta_{E,0-1}} + \sqrt{1-p_{\Delta_{0-1}}}\ket{1_q}\ket{0_E}\\
    U\ket{2_q}\ket{0_E} = \sqrt{p_{\Delta_{1-2}}}\ket{1_q}\ket{\Delta_{E,1-2}} + \sqrt{1-p_{\Delta_{1-2}}}\ket{2_q}\ket{0_E}
\end{eqnarray}

where \textit{q} labels the qutrit, and $E$ labels the environment. $\ket{0_E}$ represents the vacuum state of the environment and $\ket{\Delta_{E,0-1}}$ and $\ket{\Delta_{E,1-2}}$ label states of the environment with energies corresponding to the $1 \rightarrow 0$ and $2 \rightarrow 1$  transitions of the qutrit respectively.\\

Note that we are ignoring the possibility of a $2 \rightarrow 0$ transition since a two-photon transition is highly suppressed in real transmon devices. At a finite temperature, there would also be a possibility of spontaneous excitation, resulting in generalized amplitude damping. However, here we disregard this possibility since the gap between the qutrit levels is much larger than $k_BT$, where $T$ is the operating temperature of the qutrit.\\ 

By tracing out the environment, we can derive the quantum channel acting on the qutrit as a result of the amplitude damping:\\

\begin{equation}
    \mathcal{E}_{AD}(\rho) = Tr_E(U(\rho \otimes \ket{0_E}\bra{0_E})U^\dagger)) = \sum_i K_i \rho K^\dagger_i
\end{equation}
where we have assumed that the environment starts in the $\ket{0_E}$ state and the Kraus operators are defined as $K_i = \langle i_E\lvert U\rvert 0_E \rangle$. Since a three-level environment suffices to model this interaction, we end up with three Kraus operators, whose matrix elements can be read off from the definition of $U$ above:

\[
\begin{array}{ccc}
\begin{aligned}
K_0 &= \begin{bmatrix}
1 & 0 & 0 \\
0 & \sqrt{1-p_{\Delta_{0-1}}} & 0\\
0 & 0 &\sqrt{1-p_{\Delta_{1-2}}} \\
\end{bmatrix}
\end{aligned}, &
\begin{aligned}
K_1 &= \begin{bmatrix}
0 & \sqrt{p_{\Delta_{0-1}}} & 0 \\
0 & 0 & 0 \\
0 & 0 & 0\\
\end{bmatrix},
\end{aligned} &
\begin{aligned}
K_2 &= \begin{bmatrix}
0 & 0 & 0 \\
0 &0& \sqrt{p_{\Delta_{0-1}}}  \\
0 & 0 & 0\\
\end{bmatrix}
\end{aligned}
\end{array}
\]

\subsection{Dephasing channel}
\label{dephasing}
In order to suppress the off-diagonal elements of the density matrix without affecting the diagonal entries, we can probabilistically apply a phase flip. To see this, observe that $Z_{12}\rho Z_{12}^\dagger $ adds a negative sign to some of the off-diagonal elements but leaves the diagonal invariant. Note that since $Z_{01}$ and $Z_{12}$ are sufficient to suppress all off-diagonal terms and $Z_{02}$ is a second order effect, we will only consider the former two. Therefore, we can model qutrit dephasing with the following map:  \\
\begin{equation}
    \mathcal{E}_{D}(\rho) = p_{0-1}Z_{01}\rho Z_{01}^\dagger + p_{1-2}Z_{12}\rho Z_{12}^\dagger + (1-p_{0-1}- p_{1-2}) \rho
\end{equation}
\subsection{Relating noise channels to $T_1$ and $T_2$ times}
We model the noise in a qutrit as a composition of amplitude damping and dephasing channels: $ \mathcal{E}_D(\mathcal{E}_{AD}(\rho))$. Note that the resulting state has 3 × 3 = 9 Kraus operators and 4 free parameters, which we must characterize from experiments. \\

To characterize $p_{\Delta_{0-1}}$, we imagine running the following experiment: 
\begin{enumerate}
    \item  Initialize the qutrit to $\ket{0_q}$
    \item Apply $X_{01}$ to excite qutrit to $\ket{1_q}$
    \item  Apply noise channel $k$ times
    \item Measure the probability of finding the qutrit in $\ket{1_q}$
\end{enumerate}

We can analytically work out this probability to be equal to: $(1 - p_{\Delta_{0-1}}) ^k$. Setting this equal to the measured probability would let us solve for $p_{\Delta_{0-1}}$. However, note that in experiments we do not have the ability to apply the noise model for a discrete number of times. Instead, we expose the qutrit to noise for a range of durations and find that the probability of measuring it in $\ket{1_q}$ afterward decays exponentially. Therefore, we set: $(1 -p_{\Delta_{0-1}})$ = $e^{-T/T_1}$, where $T_1$ is the decay rate measured from experiments. Similarly, we can characterize $ p_{\Delta_{1-2}}$ by studying the decay of the qutrit from $\ket{2_q}$ to $\ket{1_q}$. \\ 

 To characterise $p_{0-1}$ we run the $T_2$ echo experiment: 
 \begin{enumerate}
     \item Initialize the qutrit to $\ket{0_q}$
     \item Apply $\sqrt{X_{01}}$ 
     \item Apply noise channel $k$ times
     \item Apply ${X_{01}}$
     \item Apply noise channel $k$ times
     \item Apply $\sqrt{X_{01}}$ 
     \item Measure the probability of finding the qutrit in $\ket{0_q}$
 \end{enumerate}
 
This probability can be found analytically to be equal to: $\frac{1}{2} (1 + (1 - 2p_{0-1})^{ 2k}(1 - p_{\Delta_{0-1}})^k)$. Taking the continuous limit, we set: $(1 - 2p_{{0-1}})$ = $e^{-T/T_2}$  . The equivalent experiment for $p_{1-2}$ involves replacing $\ket{0_q}$ with $\ket{1_q}$ and the 0-1 gates with their 1-2 equivalents.\\

The value of $T_1$ and $T_2$ is different for each qutrit on the device. For the purposes of our noise model, we pick typical values of 100 $\mu$s and 75 $\mu$s for $T_1$ and $T_2$ respectively, for both $0-1$ and $1-2$ levels. The gate durations of the CNOT gate and native single qutrit gate were chosen to be 500 ns and 40 ns, respectively. 

\subsection{Dynamical decoupling}

\begin{figure}[htbp!]
\begin{center}
\includegraphics[width=0.5\linewidth]{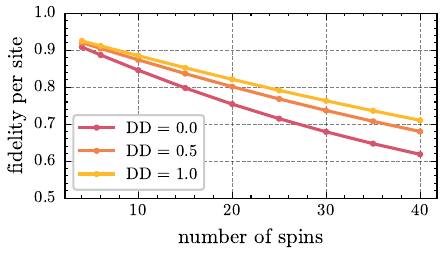}
\end{center}
\caption{\textbf{Effect of dynamical decoupling on state preparation} The ladder structure of our state preparation circuit makes it amenable to noise reduction via dynamical decoupling. Fidelity per site is defined as the $n$th root of the two-norm fidelity, where $n$ is the system size. DD = 0.0 corresponds to no dynamical decoupling, while a DD of 1.0 corresponds to perfect dynamical decoupling with effective $T_2$ time equal to $2T_1$.} 
\label{dd_figure}
\end{figure}

We note here that an advantage of the state preparation scheme described in Supplementary Material \ref{circuit_compilation} is the ladder structure of the circuit. Since each qudit remains idle for a significant portion of the circuit, we can take advantage of dynamical decoupling \cite{dd} schemes to mitigate the effect of noise. Dynamical decoupling corresponds to turning off the dephasing component of the noise channel, leading to an effective $T_2$ time between the measured $T_2$ and $2T_1$, with the latter corresponding to perfect decoupling. In Figure \ref{dd_figure}, we simulate the effect of dynamical decoupling on the fidelity per site, defined as the $n$th root of the two-norm fidelity, where $n$ is the system size. We see that the ladder structure of the ansatz allows signficant improvement from the implementation of dynamical decoupling. 

\subsection{Tensor network simulation of gate-based noise}

In order to simulate the effect of noise in a quantum circuit, we must promote each gate, $V$, in the circuit to a superoperator:
\begin{equation}
    \Lambda_V = \left(\sum_i K_i^* \otimes K_i \right) \circ  \left(V^* \otimes V \right)
\end{equation}
where $K_i$ are Kraus operators that depend on measured $T_1$, $T_2$ times and the gate time $T$. Note that it is helpful to think of $\Lambda_V$ as a tensor with four indices. For a qutrit, the dimension of each index is three. Our simulator computes the action of this superoperator on a state by contracting two of its indices to those of a density matrix represented as a matrix product operator. 

\section{Hadamard test for states prepared using post-selection}
\label{gen_hadamard_test}
\begin{figure}[htbp!]
    \begin{tikzpicture}
        \node{\begin{quantikz}[row sep = 0.8 cm,column sep =0.5
 cm]
\lstick{\text{\textit{h}:} $\ket{0}$} & \gate[style={fill=green!10}]{H_{01}} &\gate[style={fill=green!30}]{S_{01}^\dagger} \gategroup[1,steps=1, style={dashed, rounded corners},background, label style ={label position = above, anchor=north, yshift=0.3 cm}]{{
for imaginary part}}& \ctrl{1}&\ctrl{2}&\gate[style={fill=green!40}]{X_{01}}\gategroup[2,steps=3, style={dashed, rounded corners},background, label style ={label position = above, anchor=north, yshift=0.3 cm}]{{
if $x=1$}}&\ctrl{1}&\gate[style={fill=green!40}]{X_{01}} && \gate[style={fill=green!40}]{X_{01}} \gategroup[4,steps=3, style={dashed, rounded corners},background, label style ={label position = above, anchor=north, yshift=0.3 cm}]{{
if $y=1$}}& \ctrl{3} & \gate[style={fill=green!40}]{X_{01}} & \gate[style={fill=green!10}]{H_{01}} & \meter{}\\
\lstick{\text{\textit{a}:} $\ket{0}$}&&&\gate[2,style={fill=purple!40}]{U_{\phi}}&&&\targ{}&&&&&&&\meter{}\\
\lstick{\text{\textit{s}:} $\ket{0}$}&&&&\gate[2,style={fill=purple!40}]{U_{\psi}^\dagger}&&&&&&&&&\\
\lstick{\text{\textit{b}:$\ket{0}$}}&\gate[style={fill=green!40}]{X_{01}}\gategroup[1,steps=1, style={dashed, rounded corners},background, label style ={label position = below, anchor=north, yshift=-0.2 cm}]{{
if $y=1$}}&&&&&&&&&\targ{}&&&
\end{quantikz}
};
    \end{tikzpicture}\\  
     \caption{\textbf{Hadamard test for states prepared using post-selection}. The real and imaginary parts of $\langle \psi|y_b \rangle \langle x_a|\phi \rangle$ can be computed by estimating measurement probabilities of the $h$ and $a$ registers.}  \label{hadamard_ps}
\end{figure}
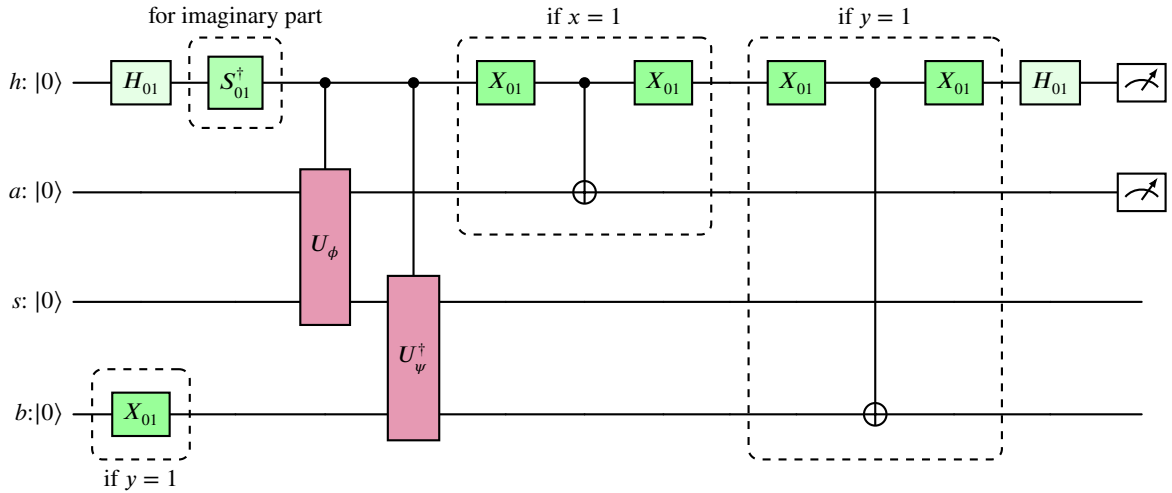
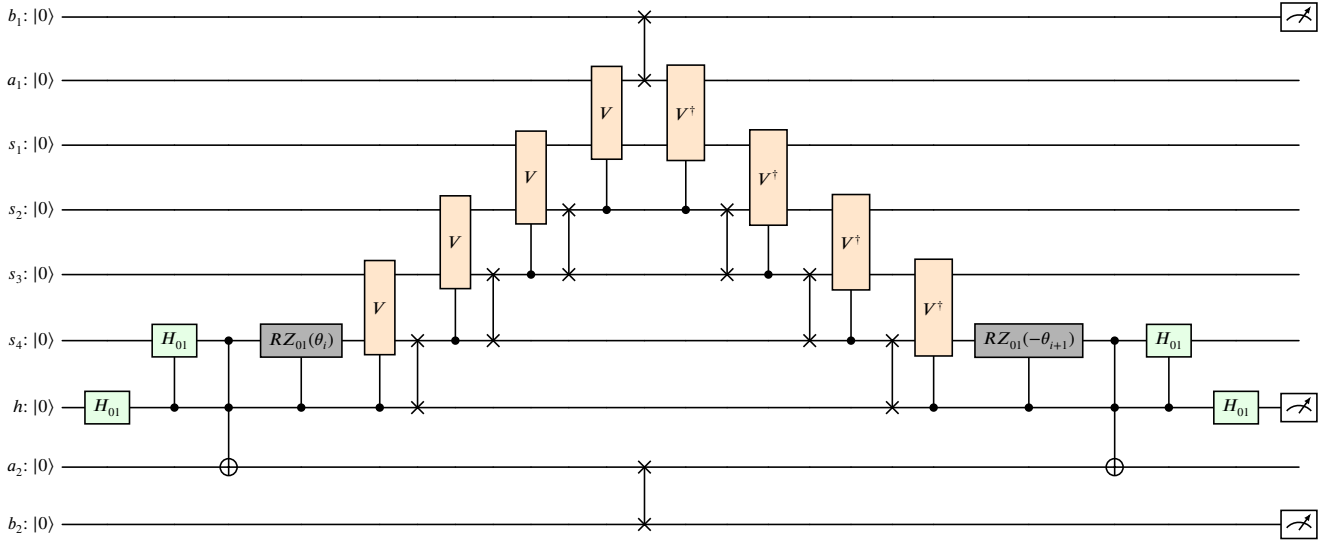
\begin{figure}[htbp!]
    \begin{tikzpicture}
        \node[scale =0.75]{\begin{quantikz}[row sep = 0.6 cm,column sep =0.4
 cm]
\lstick{\text{\textit{$b_1$}:} $\ket{0}$} &&& &&&&&&&&&\swap{1}&&&&&&&&&&&&\meter{}\\
\lstick{\text{\textit{$a_1$}:} $\ket{0}$} && &&&&&&&&&\gate[2, style={fill=orange!20}]{V}&\targX{}&\gate[2, style={fill=orange!20}]{V^\dagger}&&&&&&&&&&&\\
\lstick{\text{\textit{$s_1$}:} $\ket{0}$} &&& &&&&&&\gate[2, style={fill=orange!20}]{V}&&&&&&\gate[2, style={fill=orange!20}]{V^\dagger}&&&&&&&&&\\
\lstick{\text{\textit{$s_2$}:} $\ket{0}$} && &&&&&\gate[2, style={fill=orange!20}]{V}&&&\swap{1}&\ctrl{-1}&&\ctrl{-1}&\swap{1}&&&\gate[2, style={fill=orange!20}]{V^\dagger}&&&&&&&\\
\lstick{\text{\textit{$s_3$}:} $\ket{0}$} & &&&&\gate[2, style={fill=orange!20}]{V}&&&\swap{1}&\ctrl{-1}&\targX{}&&&&\targX{}&\ctrl{-1}&\swap{1}&&&\gate[2, style={fill=orange!20}]{V^\dagger}&&&&&\\
\lstick{\text{\textit{$s_4$}:} $\ket{0}$} & &\gate[style={fill=green!10}]{H_{01}}& \ctrl{1}&\gate[style={fill=black!30}]{RZ_{01}(\theta_i)}&&\swap{1}&\ctrl{-1}&\targX{}&&&&&&&&\targX{}&\ctrl{-1}&\swap{1}&&\gate[style={fill=black!30}]{RZ_{01}(-\theta_{i+1})}&\ctrl{1}&\gate[style={fill=green!10}]{H_{01}}&&\\
\lstick{\text{\textit{$h$}:} $\ket{0}$} & \gate[style={fill=green!10}]{H_{01}}&\ctrl{-1}&\ctrl{1}&\ctrl{-1}&\ctrl{-1}&\targX{}&&&&&&&&&&&&\targX{}&\ctrl{-1}&\ctrl{-1}&\ctrl{1}&\ctrl{-1}&\gate[style={fill=green!10}]{H_{01}}& \meter{}\\
\lstick{\text{\textit{$a_2$}:} $\ket{0}$} & &&\targ{}&&&&&&&&&\swap{1}&&&&&&&&&\targ{}&&&\\
\lstick{\text{\textit{$b_2$}:} $\ket{0}$} & &&&&&&&&&&&\targX{}&&&&&&&&&&&&\meter{}\\
\end{quantikz}
};
    \end{tikzpicture}\\  
     \caption{\textbf{Routing of Hadamard register}: Hadamard test circuit for $\langle GS(\theta_{i+1})|0_b \rangle \langle 0_a|GS(\theta_i) \rangle$, where $GS(\theta_i)$ represents the ground state of the PBC AKLT model with perturbation angle, $\theta_i$. By exploiting the ladder structure in our state preparation protocol, we find an efficient routing strategy for the Hadamard register that minimizes the number of SWAP gates necessary. }  \label{hadamard_detailed}
\end{figure}

In this section, we describe a generalized Hadamard test that can be used to compute the overlap of two states prepared using post-selection. Suppose we want to compute the overlap of $\ket{\phi_t}$ and $\ket{\psi_t}$ and have circuits $U_{\phi}$ and $U_{\psi}$ such that: 
\begin{align}
\ket{\phi_t} &= \frac{1}{\sqrt{p_0}}\langle 0_a|\phi \rangle + \frac{1}{\sqrt{p_1}}\langle 1_a|\phi \rangle = \frac{1}{\sqrt{p_0}}\bra{0_a}U_{\phi}\ket{0_s}\ket{0_a} + \frac{1}{\sqrt{p_1}}\bra{1_a}U_{\phi}\ket{0_s}\ket{0_a} \\ 
\ket{\psi_t} &= \frac{1}{\sqrt{q_0}}\langle 0_b|\psi \rangle + \frac{1}{\sqrt{q_1}}\langle 1_b|\psi \rangle = \frac{1}{\sqrt{q_0}}\bra{0_b}U_{\psi}\ket{0_s}\ket{0_b} + \frac{1}{\sqrt{q_1}}\bra{1_b}U_{\psi}\ket{0_s}\ket{0_b},
\end{align}
where the $s$ labels system register, $a,b$ label ancilla registers, and $\ket{0}$ and $\ket{1}$ represent the all zero and all one state respectively on the labeled register. Note that while the system register is shared between the two states, each state has its own ancilla register. The probabilities of measuring the ancilla register of $\ket{\phi}$ in the all zero and all one state are represented by $p_0$ and $p_1$, respectively. The equivalent quantities for $\ket{\psi}$ are represented by $q_0,q_1$. \\

The overlap between the two states can be written as: 
\begin{align}
    \langle \psi_t| \phi_t \rangle &= \frac{1}{\sqrt{p_0q_0}}\langle \psi|0_b \rangle \langle 0_a|\phi \rangle + \frac{1}{\sqrt{p_0q_1}}\langle \psi|1_b \rangle \langle 0_a|\phi \rangle + \frac{1}{\sqrt{p_1q_0}}\langle \psi|0_b \rangle \langle 1_a|\phi \rangle + \frac{1}{\sqrt{p_1q_1}}\langle \psi|1_b \rangle \langle 1_a|\phi \rangle 
\end{align}

The circuit in Fig. \ref{hadamard_ps} allows us to compute the real and imaginary parts of each of these four terms. Specifically, we can relate each term to certain measurement probabilities of an appropriate circuit. For instance, the circuits used to compute the real part have the following measurement probabilities:
\begin{align}
    \text{prob}(h=0, a=x) &= \frac{1}{4}(1 + p_x + 2\text{Re} \langle\psi|y_b\rangle \langle x_a \phi\rangle)
\end{align}
where $x,y\in [0,1]$. Therefore, estimating these probabilities will allow us to extract the complex overlap. \\

We can use the modified Hadamard test to compute the Berry phase of the PBC AKLT state prepared as described in Section \ref{scalable_ansatz} in the main text. Note that for generic states, the controlled-$U$ gates depicted in Fig. \ref{hadamard_ps} will turn into long-range gates, requiring a large number of SWAP gates. However, the ladder structure in Fig. \ref{scalable_fidelity} A in the main text leads to an efficient routing of the Hadamard register and drastically reduces the number of SWAP gates necessary. This routing strategy is depicted in Fig. \ref{hadamard_detailed}. The long-range CNOT gates at the end of the Hadamard circuits can also be efficiently implemented in cases where the size of the circuit matches the length of a closed loop of qutrits on a two-dimensional device. This is because the CNOT gates only stretch across the edge qutrits and, therefore, we can exploit periodicity to decompose the long-range CNOTs into local CNOTs and SWAP gates that traverse only the edge of the circuit. 

The gates necessary to implement these circuits can be compiled using the method described in Section \ref{circuit_compilation}.

\end{document}